# Structural, Microstructural and Electrochemical Properties of Dispersed Type Polymer Nanocomposite Films

**Anil Arya and A. L. Sharma***

Centre for Physical Sciences, Central University of Punjab, Bathinda-151001, Punjab, India

E-Mail address:alsharmaiitkgp@gmail.com

**Abstract**

The free standing solid polymer nanocomposite (PEO–PVC)+$LiPF_6$-$TiO_2$ films has been prepared through standard solution cast technique. The improvement in structural, microstructural and electrochemical properties has been observed on the dispersion of nanofiller in polymer salt complex. X-Ray diffraction studies clearly reflect the formation of complex formation as no corresponding salt peak appeared in the diffractograms. The FTIR analysis suggested a clear and convincing evidence of polymer–ion, ion-ion and polymer-ion-nanofiller interaction. The highest ionic conductivity of the prepared solid polymer electrolyte films is ~$5\times10^{-5}$ S $cm^{-1}$ for 7 wt. % $TiO_2$. The Linear Sweep Voltammetry provide the electrochemical stability window of the prepared solid polymer electrolyte films, which is of the order of ~3.5 V. The ion transference number has been estimated, $t_{ion}$ =0.99 through dc polarization technique. Dielectric spectroscopic studies were performed to understand the ion transport process in polymer electrolytes. All solid polymer electrolyte possesses good thermal stability up to 300 °C. DSC analysis confirms the decrease of melting temperature and signal of glass transition temperature with addition of nanofiller which indicates the decrease of crystallinity of polymer matrix. An absolute correlation between diffusion coefficient (D), ion mobility (μ), number density (n), double layer capacitance ($C_{dl}$), glass transition temperature, melting temperature ($T_m$), free ion area (%) and conductivity (σ) has been observed. A convincing model to study the role of nanofiller in polymer salt complex has been proposed which supports the experimental findings. The prepared polymer electrolyte system with significant ionic conductivity, high ionic transference number, good thermal, voltage stability could be suggested as a potential candidate as electrolyte cum separator for fabrication of rechargeable lithium-ion battery system.

Keywords: polymer blend, structural properties, glass transition temperature, transport parameters, transport mechanism

## 1. Introduction

From last two decades, lithium ion batteries (LIB) are widely investigated energy source for the portable electronics because of attractive features like high energy density, shape mouldability, design flexibility, zero memory loss and high safety. The desirable characteristics of polymer electrolytes for R&D application in LIB are good compatibility with electrode materials, leakage proof, longer shelf life, high capacity and light weight [1]. The conductivity in solid state conductors was reported first time by Faraday in 1800s, and later the discovery of poly ethylene oxide (PEO) and alkali metal salts (NaI) system in the 1970s was contributed by Wright and coworkers. Then, Armand realized their technological importance due to flexibility, deformability and introduced the first new class of solid state ionic conductors (SSIC). Solid Polymer electrolytes (SPEs) comprising the high electrical conductivity along with mechanical flexibility, faster segmental motion of polymer chain makes them, best substitute over liquid electrolytes in electrochemical devices such as supercapacitor, cell phones, lithium ion batteries, fuel cells, and smartphones. The liquid electrolytes offer limitations such as chemical reactivity with lithium metal electrode, leakage, and need of additional inert separator for device applications. While, poor mechanical stability, thermal stability and chemical stability of gelled polymer electrolytes (GPE) due to the presence of polar solvents imparting semi-solid/liquid behavior bounds of their use in storage/conversion devices [2-5].

The SPE act as both the separator and the electrolyte, with the aim of improving battery's design and reliability [6-7]. SPEs usually consist of an amorphous high molecular weight polymer matrix and salt having large anion dissolved in it to reduce ion aggregation. The polymer contains ligand coordination to the cations of the salt, and provides the key solvation enthalpy to promote the formation of the polymer electrolyte. The polymer chains consistently create suitable coordinating sites adjacent to the ions; facilitate the movement of ions via segmental motion. The primary focus of the present study is to improve the ionic conductivity of SPEs which could be achieved by two means (1) improvement in polymer chain mobility by suppressing crystallization polymer chains; (2) increase in a number of free ion charge carriers. The improvement in the polymer chain mobility can be obtained by cross-linking and blending, out of two possibilities blending one is most attractive, because blending exhibits superior properties than an individual component of the blend. Polymer blends are obtained by physical mixing of two or more polymers without any chemical reaction between them. This gives in hand superior properties than individual polymer and can be easily controlled by varying the composition of polymer used [8-10].

Polymer blending technique is the best approach to develop the new polymeric materials for application prospects. Recently, many efforts have been devoted to enhance the electrochemical and mechanical properties in blend polymer electrolyte membranes such as PEO-PAN [11], PVdF-PEO [12], PEO-PVdF [13], PVA-PEO [14], PEO-PEG [15], PVA-PVP [16], PVC-PEMA [17], PEO-PVP [18] and PVC-PEO [19-21]. Various strategies are considered for enhancing the ionic conductivity of SPEs, such as; the addition of plasticizers, organic liquids, nanofiller and organic clay. To achieve the high ionic conductivity and mechanical properties doping of nano-sized fillers/plasticizers having large surface area is an effective approach. This also increases the amorphous content which results in faster transport of ions by coordinating sites of the host polymer. The addition of nanofiller with high dielectric constant and high thermal stability improves the mechanical, thermal, and electrochemical properties of electrolyte due to increase in free volume and leads to increased mobility of polymer chain and reduction in crystallinity [22-34].

To date, the high molecular weight Poly (ethylene oxide) (PEO)-based polymeric electrolytes are researched widely due to their semi crystalline nature and presence of ether group which supports faster ionic transport due to beneficial polymer flexibility [35, 36]. The host polymer is polyethylene oxide (PEO) containing ether group (-Ö-) which provides coordinating sites to cation ($Li^+$) and $Li^+$ acts as Lewis acid. Hence it can form complex easily with many alkalis salts and provides a direct path for cation migration due to the presence of ether group $(–CH_2–CH_2–\ddot{O}–)_n$ in the polymer backbone. However, in PEO presence of C-C, C-O and C-H bonds low reactivity offer them better chemical and electrochemical stability, but poor ionic conductivity and thermal stability [18]. To overcome the limitation of lower ionic conductivity and limited working temperature (100 $^oC$) in PEO due to low melting temperature (65 $^oC$), Polyvinyl chloride (PVC) has been chosen as partner to PEO for synthesizing blend polymer electrolyte which permits the faster ion migration than the semi-crystalline polymer. supports the superior mechanical and thermal properties of the other polymers. PVC ($(C_2H_3Cl)_n$) has an amorphous structure, resistant to oxidative reactions, high melting temperature (∼260 $^oC$) and acts as an excellent mechanical stiffener. Also blending of PVC with PEO provides a wide temperature range of operation (∼200 $^oC$) as compared to pure PEO which is a vital requirement for energy storage devices [37]. The salt is chosen due to its unique properties such as good ionic conductivity, the smaller cation size (~0.76 Å), the bulky anion (~ 2 Å) and low lattice energy. Other salts with a large anion such as LiTFSI and LiBETI are also available, but corrosion with aluminum collector limits their use in commercial devices [38]. So, $LiPF_6$ is best among all salt as it provides the adequate balance of electrical, mechanical and thermal properties. The Lewis acid group of inert nanofiller (Ti+2) may compete with Lewis acid Li+ for the formation of a complex with salt. These interactions result in the lowering the ion coupling, by structural modification and promote more salt dissociation for free ions. As polymer phase is the medium of transport of ions and nanofiller interaction with polymer phase affects polymer chain stiffness. Also, there may be the direct interaction of nanofiller with PEO, cation, and anion due to a higher dielectric constant of TiO2 ($\varepsilon$=85) than PEO ($\varepsilon \sim 5$) [39]. Figure 1 displays the molecular structures of the materials used in present study.

Among the nano-sized ceramic fillers explored recently, $TiO_2$ seems to be the most promising candidate to increase the ionic conductivity due to its high Lewis-acid character as compared to other commonly used ceramic fillers (such as $Al_2O_3$, $SiO_2$, and $ZrO_2$). Scrosati et al., [27] also studied the PEO–$LiClO_4$ complexes with $TiO_2$ and $Al_2O_3$ ceramic fillers and reported that $TiO_2$ was the best filler which exhibited the greatest enhancement in ionic conductivity. Chung et al., [40] studied the effects of $TiO_2$, $Al_2O_3$, and $SiO_2$ in PEO–$LiClO_4$ polymer electrolyte and identified $TiO_2$ as the filler with the greatest enhancement in ionic conductivity due to the weak interactions between the polymer chain and the $Li^+$ ions. Although, it has been compared by taking different nano-filler of same concentration with PS complex and found that the conductivity maxima appears to be for $Er_2O_3$, but sample preparation for low concentration of $Er_2O_3$ are not so easy therefore selection of $TiO_2$ as nano-fillers seems to be more appropriate for better study at varying concentrations and in view of optimization of nano-filler concentrations [41]. Vignarooban et al., [42] reported the enhancement of ionic conductivity of poly (ethylene oxide)–lithium trifluoromethanesulfonate ($LiCF_3SO_3$ or LiTf) based polymer electrolyte by incorporating $TiO_2$ nano-filler. To the best of our knowledge, there is no study available in the literature on the ionic conductivity enhancement of PEO–PVC+$LiPF_6$ based solid polymer electrolyte on the addition of nano-sized $TiO_2$ ceramic filler. Contextually, this is the first report with $LiPF_6$ salt for PEO-PVC blend based solid polymer electrolyte. The present work discusses the preparation of SPE with different nanofiller content by universal solution cast technique. The structural, thermal, electrical and transport properties of polymer nano-composites have been investigated. Finally, a self- proposed ion transport mechanism is established to support experimental data.

## 2. Experimental

### *2.1. Materials*

All the solid polymer electrolyte (SPEs) samples were prepared by a standard solution cast technique. The SPEs system involved the use of poly (ethylene oxide) (PEO) and poly (vinyl chloride) (PVC) with average molecular weights of $1\times10^6$ (Sigma-Aldrich), and $6\times10^4$ respectively. Salt $LiPF_6$ and Titanium oxide ($TiO_2$) were acquired from Sigma-Aldrich. Anhydrous tetrahydrofuran (THF) was used as solvent purchased from Sigma-Aldrich.

### *2.2. Preparation of solid polymer electrolyte*

All the Solid Polymer Electrolyte (SPEs) films were prepared by a standard solution cast technique reported elsewhere [7]. A snapshot of the method is shown in the flow chart/diagram (Figure 2). The polyethylene oxide (PEO) was stirred in tetrahydrofuran (15 ml) for 8 hr. and then polyvinyl chloride (PVC) was added to the solution and again stirred to carry out physical blending for 10 hr. Subsequently, an optimized amount of (stoichiometric) $LiPF_6$ (Ö/Li= 8) was added to the polymer blend and stirred mechanically for 12 h to allow homogenous mixing and complexation. Next, $TiO_2$ nanofiller was added to the polymer salt complex solution in different $x$ wt. % concentration ($0 \leq x \leq 20$; $x$ refers to nanofiller wt. % concentration) and stirred for 12 h. Then obtained solution was sonicated for 30 minutes for homogenous mixing of nanofiller. The prepared polymer nanocomposite solutions were casted in petri dishes and kept at 60 °C for 24 hours in vacuum oven. Finally, the free standing polymer films were peeled off the petri dishes and stored in vacuum desiccator with silica gel to avoid any moisture content.

*2.3. Characterization*

X-ray diffraction (XRD) is the most useful method for investigating the structural properties and crystallinity of the polymer matrix. The X-ray diffraction was recorded with Cu-$K_\alpha$ radiation (λ=1.54 Å) in the Braggs angle range (2θ) from 10° to 60° with a scan rate of 2° per minute. Field emission scanning electron microscopy (FESEM) was used to study the surface morphology (FESEM: Carl Zeiss product) and taken in a high vacuum after sputtering the samples with gold to prepare conductive surfaces. The Fourier transforms infrared (FTIR) spectra were recorded using the Bruker Tensor 27 (Model: NEXUS –870) in absorbance mode over the wavenumber region from 600 to 4000 $cm^{-1}$ with a resolution of 4 $cm^{-1}$. FTIR is performed to probe the presence of microscopic interactions between polymer-ion, ion-ion and polymer-ion-nanofiller. The electrical conductivity of the solid polymer electrolyte films was evaluated out using Electrochemical Analyzer (CHI 760; USA) over the frequency range from 1 Hz to 1 MHz at a signal level of 10 mV by sandwiching SPE between two stainless steel (SS) electrodes at room temperature. The intercept between the semi-circle at high frequency and tilted spike at low frequency were taken as the bulk resistance ($R_b$). The ionic conductivity (σ) value was obtained using equation 1:

$$\sigma_{dc} = \frac{1}{R_b}\frac{t}{A} \qquad (1)$$

Where t is thickness (cm) of polymer film, $R_b$ is bulk resistance (Ω) and A is area ($cm^2$) of working electrode. The real (Z′) and imaginary (Z″) parts of the complex impedance were used to calculate the real (ε′), imaginary (ε″) parts of permittivity and ac conductivity ($\sigma_{ac}$).The total ionic transference number ($t_{ion}$) was obtained by sandwiching polymer electrolyte film between stainless steel (SS) blocking electrodes and a fixed dc volateg of 0.05 V was applied across the $SS|SPE|SS$ cell. Ion transference numbers of the solid polymer electrolytes was evaluated using equation 2 on $SS|SPE|SS$ cell:

$$t_{ion} = \left(\frac{I_t - I_e}{I_t}\right) \times 100 \qquad (2)$$

$I_t$ and $I_e$ are the total and residual current respectively.

The linear sweep voltammetrey (LSV) was characterized to obtain the working volatge window of the electrolyte. Thermal stability of the synthesized SPE films was investigated using thermo-gravimetric analysis (TGA – SHIMDZU–DTG-60H) under dynamic temperature conditions from 30 °C to 600 °C, in a controlled nitrogen atmosphere at a constant heating scan rate of 10 °C $min^{-1}$. Differential scanning calorimetry (DSC) measurements were performed to find the glass transition temperature and crystallnity of all SPEs with a heating rate of 10 ˚C $min^{-1}$ from -100 to 100 ºC under $N_2$/Ar atmosphere (DSC-Sirius 3500). SPEs films with the weight of 8-10 mg were sealed in aluminum pans, and an empty aluminum pan was used as a reference.

## 3. Results

### 3.1. X-ray diffraction (XRD) analysis

Figure 3 represents the XRD patterns of a PEO-PVC polymer blend, blend polymer salt complex with different concentrations of $TiO_2$. XRD pattern of PEO-PVC film exhibits two partially crystalline characteristic diffraction peaks at 2θ =19.30° and 23.47° for PEO relevant to [120] and [032]+[112] planes [43]. The above inspection shows the presence of multiphase character consisting of both crystalline and amorphous nature. In contrast, the addition of salt ($LiPF_6$) to the polymer blend shows a shift in peaks toward lower angle side 18.67° and 23.26°.  Also, the peak intensity decreases which suggest the reduction in the crystallinity of the polymer blend [44]. It may be due to the strong interaction of cation ($Li^+$) with electron rich ether group of polymer host. It concludes that amorphous phase is enhanced by accommodation of the lithium ion ($Li^+$) in the polymer backbone. Above analysis is also in correlation with correlation given by Hodge et al, [45] between the degree of crystallinity and peak intensity, which states that higher crystallinity is associated with greater peak intensity. The absence of peak corresponding to $LiPF_6$ reveals complete dissociation of the salt in the blend polymer electrolyte (BPE). Besides these crystalline peaks, the broad hump near 25° is also observed, and it arises from the amorphous phase of the PVC [46-47]. The sharpness of diffraction peaks corresponding to PVC is same in all samples which suggest that salt interacts with PEO and amorphous phase formation occurs due to disruption of crystalline chain arrangement. The addition of nanofiller has brought noticeable changes in the peaks with a shift in position and broadening. The XRD pattern of $TiO_2$ added SPE (Fig 3 c-h) exhibits lower intensity peaks located at 26.70°, 36.80°, 45.05°, 54.8°, respective to [101], [112], [200] and [211] planes [44].The strong diffraction peaks at 2θ =25.4°, 48° confirms the $TiO_2$ anatase structure and absence of broad diffraction peak on the addition of nanofiller evidence the modification in the polymer salt matrix.

The d-spacing between the diffraction planes was obtained using the Bragg's formula $2d\sin\theta=n\lambda$, crystallite size L of PEO crystallines from Sherrer's equation $L=0.94\lambda/\beta\cos\theta$ and interchain separation (R) using the equation $R=5\lambda/8\sin\theta$ and determined values are shown in Table 1. It is found that the Basal spacing, crystallite size change with the addition of the nanofiller and may be attributed to the polymer-ion-nanofiller interaction. The increase of interchain separation (R) evidences the increase of the amorphous content or free volume after addition of the nanofiller due to nanofiller interaction with the polymer chains. This increase of the free volume promotes the faster ion migration due to the faster segmental motion of the polymer chains. Addition of nanofiller makes the polymer chains more flexible by reducing the covalent binding between the polymer chains and which leads to the faster polymer chain segmental motion.

The disappearance of some small intensity peaks in polymer matrix suggests the presence of strong interaction between polymer-salt and nanofiller. The $TiO_2$ nanofiller based SPEs also displays a slight shift compared to polymer salt complex and polymer blend which is evidence of amorphous phase formation and coordinating interaction between a surface group of nanofiller and polymer host. The above coordinating interaction increases the polymer segmental motion or polymer flexibility which supports fast ion transport and directly evidences increase of ionic mobility and hence ionic conductivity.

### 3.2. Field Emission Scanning Electron Microscope (FESEM) Analysis

The electrical properties of SPEs films are completely associated with the surface morphology and a smoother, more dense film often facilitate faster transport of ions since it will be more ordered and less porous. To understand the role of salt and nanofiller on surface properties, specific efforts have been made to observe the morphologies of the blended polymer electrolyte system using FESEM.

Figure 4 depicts the field emission scanning electron micrograph (FESEM) of free-standing SPEs films. The rough interconnected surface of the salt free blend polymer (PEO-PVC) is fundamental nature of the polymer and depends on interaction as well as evaporation rate of the solvent in blend polymer. A remarkable difference in the surface roughness and texture of the polymer film on the addition of salt is evidenced as in Fig 4 b. This irregular appearance of the film confirms polymer salt complex formation. Now, the addition of even the small concentration of nanofiller improves the surface roughness, and interlining of grain boundaries are seen in the structure (Figure 4 c). The addition of nanofiller even at low wt. % concentration has modified the blend based free-standing polymer films surface morphology, and a smooth surface of the polymer with nanofiller is seen (Figure 4 e) [48]. The appeared dark regions in micrograph might be due to the continuous growth of amorphous region of the polymer blend and this region increases with the interruption of the crystallite in the presence of the $TiO_2$ nano-filler [49].The smooth surface of sample reveals that polymer salt has good compatibility with nanofiller and is linked with high amorphous content in the polymer electrolyte. The smooth and homogenous morphology of SPE films with high nanofiller content is evidence of lowering of the crystallinity due to a constraint on chain reorganization tendency and facilitate rapid ion migration. The high amorphous content enables faster ionic transport by providing an easy path for ion migration and high mobility alongwith the enhanced flexibility of the films. It is anticipated from above that the optimized morphology of the polymer film may be reflected in the enhanced electrochemical behavior of the solid polymer electrolyte in the lithium-ion battery (see the following section).

EDX mapping is an appropriate tool to study the nanofiller dispersion inside the polymer electrolyte system. FESEM/EDS mapping was also implemented to investigate the presence and distribution of the nanofiller in the polymer matrix. The distribution of Ti atoms was visualized by the EDS elemental mapping of Ti atoms. As expected, the uniform distribution was obtained from nanofiller content, and some agglomeration was observed for high nanofiller content. As can be seen in Figure 5 that polymer matrix with 7 wt. % nanofiller content in EDS mapping shows uniform/homogenous distribution over the whole surface as indicated by red color and this indicates that the $TiO_2$ particles were well dispersed in polymer matrix. All the three figures show the almost identical mapping of the Ti, which reveals that nanofiller is mixed properly in the polymer matrix. In addition, it is remarkable that the resulting $TiO_2$ added SPE membrane is semi-transparent and free standing (Figure 2).

### 3.3. Fourier Transform Infrared (FTIR) Analysis

FTIR has been proven to be highly effective for investing various interactions such as polymer ion interaction, ion-ion interaction and polymer ion filler interaction. FTIR also confirm the presence of complex formation and demonstrates interaction among different constituents of the polymer matrix. Absorption changes the frequency of stretching and deforming vibrations of these groups and band shift occurs w.r.t wavenumber, which provides direct evidence of interaction occurring in polymer matrix system by addition of nanofiller.

Figure 6 shows the FTIR absorbance spectra of (PEO-PVC)-$LIPF_6$ + $x$ wt. % $TiO_2$ ($0 \leq x \leq 20$) based system in the wavenumber region 600-3000 $cm^{-1}$ and band assignment is shown in Table 2. The blend PEO-PVC spectrum shows various absorption bands as finger print region (shown by dotted line) in the region between 600 and 1500 $cm^{-1}$. The characteristic absorption bands observed in the FTIR spectrum at the wavenumbers ~620, ~845, ~958, ~1080, ~ 1250, ~1666, ~2880 and 2914 $cm^{-1}$ are attributed to C-$Cl_{st}$, $\nu_s(PF_6^-)$, trans (C-H) wagging, t$(CH_2)_s$, $\tau(CH_2)_s$, $\upsilon$(C=O), $\nu_s(CH_2)$ and $\nu_a(CH_2)$ respectively [5, 50]. The most of the band located at 1250 $cm^{-1}$, 1320 $cm^{-1}$, 1360 $cm^{-1}$, 1398 $cm^{-1}$, 1435 $cm^{-1}$ and 1465 $cm^{-1}$ are attributed to twisting $(CH_2)_s$ , stretching $(CH_2)$, symmetric $CH_2$ wagging , asymmetric$(CH_2)_a$ wagging, deformation $(CH_2)_s$ and 1465 bending $(CH_2)_a$ respectively exhibited by PEO [36]. The $CH_2$ stretching mode also undergoes some variations which result in the decreased intensity and peak broadening in the range 2850-2950 $cm^{-1}$ with addition of nanofiller. The peaks appearing near 720 $cm^{-1}$ corresponds to the C-Cl stretching mode and at 1060 $cm^{-1}$ assigned to (C-C) stretching mode of PVC. The absorption peak located near 958 $cm^{-1}$ corresponds to cis CH wagging and trans CH wagging mode of PVC respectively. The bands present near 958 $cm^{-1}$, 1190 $cm^{-1}$ and 1254 $cm^{-1}$ are assigned to CH wagging, $CH_2$ twisting and $CH_2$ rocking mode of PVC respectively [20, 51]. A peak was observed at 1060 $cm^{-1}$ in PEO-PVC blend associated with the crystalline structure and peak intensity decreases after addition of salt (Fig 6 b). Further addition of nanofiller demonstrates that it gets disappeared indicating the reduction in crystallinity (Fig 6 c-h) [52]. In the observed spectra, there are no distinct bands corresponding to $TiO_2$ nanofiller. The characteristics peaks of pure PEO, (1114 $cm^{-1}$, 1350 $cm^{-1}$), are shifted to 1112 $cm^{-1}$, 1349 $cm^{-1}$ for PEO-PVC blend. Further the characteristic peaks of pure PVC (696 $cm^{-1}$, 1240 $cm^{-1}$), are shifted to 731 $cm^{-1}$, 1245 $cm^{-1}$ for PEO-PVC blend system and above analysis in terms of peak shift evidences the complexation of PEO-PVC blend. Any interaction of ($PF_6^-$) anion changes its symmetry from $O_h$ point group to $C_{3\upsilon}$ or $C_{2\upsilon}$and changes in absorption mode of polymer matrix are evidence in FTIR spectrum between 600 $cm^{-1}$ to 1500 $cm^{-1}$. The effect of the nanofiller on the peak pattern in terms of asymmetry and peak position has been recorded in the Table 2.

*3.3.1. Polymer-ion interaction*

The characteristic absorption peak of host polymer PEO near 1110 $cm^{-1}$ is attributed to C-O-C symmetric stretching mode involving ether oxygen of PEO and provides direct evidence of complexation owing to the interaction of ether group of the polymer chain with action (Figure 7). Table 2 shows the shift in C-O-C band occurring at 1114 $cm^{-1}$ towards lower wavenumber side to 1112 $cm^{-1}$ on the addition of salt in polymer blend matrix. This shift of peak towards lower wavenumber side is strong evidence of cation coordination ($Li^+$) at electron rich ether group of PEO. The variations in the intensity, shape, and position of the C–O–C symmetric/asymmetric stretching mode were associated with the polymer–$LiPF_6$ interactions and dependence of the vibrational peaks of C–O–C on the filler concentrations were clarified in FTIR spectra. The peak broadening and decrease in intensity with the addition of nanofiller reveal the weakening of C-O-C bond due to interaction with –C=Ö- as an active site for $Li^+$ coordination(Fig 7 c-h). Figure 7 (a–h) shows the spectral pattern of hexafluorophosphate ($PF_6^-$) group with octahedral symmetry ($O_h$) in the solid polymer electrolyte films. Any possible coordination of $PF_6^-$ group with cation usually results in lowering of the $PF_6^-$ group symmetry from $O_h$ point group to $C_{3\upsilon}$ or $C_{2\upsilon}$ depending on the nature of the coordination that $PF_6^-$ group possesses (tridentate/bidentate) with cations ($Li^+$) and other such active sites having Lewis acid character. Nanofiller addition in the polymer salt matrix may lead to the interaction of nanofiller with the anions ($PF_6^-$), and changes are evidenced regarding the shift in peak. Also, the profile of C-H stretching mode near 951 $cm^{-1}$ in polymer blend shifts to higher wavenumber side on the addition of salt, suggesting the firm interaction of the polymer with cation. The correlation with electrical conductivity and study of more interaction on the addition of nanofiller is described in later section.

*3.3.2. ion-ion interaction*

The standard internal vibrational mode of $PF_6^-$ anion in SPEs films have been observed in wavenumber region 820-860 $cm^{-1}$ and this demonstrates asymmetry in peak as clearly visible in polymer salt film (Figure 8 a). This asymmetry in $\upsilon_3(t_{1u})$ mode is the result of degeneracy ($O_h \rightarrow C_{3v}$) arising due to the simultaneous presence of more than two components i.e. free ion, ion pairs and is confirmed by ion-ion interaction. This causes a change in free anion area of the nanofiller based system as compared to the polymer salt system. So to study above degeneracy and presence of both components, deconvolution of $PF_6^-$ anion peak is done for qualitative and quantitative analysis using Peak Fit (V 4.02) software with Voigt Amp profile. The best fit is said to be measured by the correlation coefficient ($r^2$). A quantitative estimation of fraction free ion and ion pair in the deconvoluted pattern is obtained using equation 3 & 4:

$$\text{Fraction of free anion} = \frac{\text{Area of free ion peak}}{\text{Total peak area}} \quad \ldots\ldots\ldots\ldots\ldots (3)$$

And

$$\text{Fraction of ion pair} = \frac{\text{Area of ion pair peak}}{\text{Total peak area}} \quad \ldots\ldots\ldots\ldots\ldots.. (4)$$

The deconvoluted pattern of SPE films with different filler concentration is displayed in figure 8. The deconvoluted pattern appearing in PS film shows two distinct peaks one at lower wavenumber and other at high wavenumber. It is general convention that peak at lower wavenumber is assigned to free $PF_6^-$ anions which do not directly interact with the lithium cations and at higher wavenumber is due to $Li^+ - PF_6^-$ contact ion pairs [53]. In the case of PS film (Figure8 a), free ion peak is located at 835 $cm^{-1}$ and ion pair at 847 $cm^{-1}$ owing to the peak asymmetry. The deconvoluted pattern for SPEs depicts changing the profile of free ion and ion pair peak in terms of change and shift in position with the addition of nanofiller [54].

A comparison of free ion area and ion pair area from Table 3 shows relatively higher free ion area for 7 and 15 wt. %. This increase in area may be attributed to release of more free charge carriers from anion $PF_6^-$ and it shows the presence of the strong interaction of nanofiller with salt. The increase in free anion concentration results in the movement in free cation concentration which does not reflect in the mid-IR region. The decrease in free anion area at high nanofiller content may be due to trapping of cation in nanofiller clusters and blockage of conduction path provided by polymer chains [55]. A high degree of ionic dissociation is achieved for 7 and 15 wt. % nanofiller content. It suggests, that the extent of interaction of the lithium cations and the $-\ddot{O}-$groups provide more successful hopping and coordinating sites to the lithium cations, resulting in an enhancement of the number of free charge carriers. In the next section, more probable interactions with the nanofiller addition such as polymer-ion-nanofiller are elaborated.

*3.3.3. Polymer ion-nanofiller interaction*

The effect of nanofiller addition in polymer salt complex results in a change in band profile of polymer matrix. Also, the effect of nanofiller is seen on the polymer-ion and ion-ion interaction in SPE films. Figure 9 shows the FTIR spectra in wavenumber range 1240 $cm^{-1}$ to 1480 $cm^{-1}$. It concludes that the various absorption band of the complex (PEO-PVC)-$LIPF_6$ + $x$ wt. % $TiO_2$ ($0\leq x \leq 20$) are influenced on addition of nanofiller $TiO_2$.

The main band present at 1280, 1350, 1434and 1467 $cm^{-1}$ are attributed to twisting $(CH_2)_s$, symmetric $CH_2$, wagging , deformation $(CH_2)_s$ and bending $(CH_2)_a$ respectively. Figure 9 shows variation in $\tau(CH_2)_s$ twisting mode of PEO with nanofiller located at 1280 $cm^{-1}$. The change in vibrational bands on the addition of nanofiller shows that the nanofiller plays an effective role in polymer salt matrix and interaction occurring between polymer ion and polymer ion interaction are evidenced [36, 56]. The intense $CH_2$ deformation at 1434 $cm^{-1}$ for blend PEO-PVC shifted to 1434 $cm^{-1}$ on addition of salt and after addition of filler from ($0\leq x\leq 20$) shift in peak is 1426 $cm^{-1}$, 1429 $cm^{-1}$, 1432 $cm^{-1}$, 1434$cm^{-1}$, 1435 $cm^{-1}$, 1438 $cm^{-1}$and 1429 $cm^{-1}$ respectively [4]. This change in wavenumber results in a change in bond length and confirms the complex formation in solid polymer electrolyte films [57, 58]. The $CH_2$ wagging mode of PEO located at 1350 $cm^{-1}$ is shifted to 1349 $cm^{-1}$ on the addition of salt. The effect of nanofiller was observed on vibration modes in FTIR spectra in terms of decrease in intensity of absorption peaks and broadening with shifting in band profile. The change in the frequency, broadening and intensity depends on the interactions of the polymer matrix with filler and surrounding in polymer nanocomposite films [51].

Another interesting IR frequency region is from 2800 $cm^{-1}$ to 3000 $cm^{-1}$ which correspond to the symmetric and asymmetric stretching mode of $CH_2$. The single peak is observed in figure 10 and presence of asymmetry confirms the presence of two peaks in this region. This is deconvoluted into two peaks at 2885 $cm^{-1}$ and 2910 $cm^{-1}$ shown in Figure 10. The former arises from the symmetric stretching ($\nu$ $(CH_2)_s$) mode, and later one is due to asymmetric stretching ($\nu(CH_2)_a$) mode . However, band splitting is obtained (figure 10) into two bands one near 2880 $cm^{-1}$ and another near 2910 $cm^{-1}$ corresponding to symmetric $CH_2$ stretching ($\nu(CH_2)_s$), asymmetric $CH_2$ stretching ($\nu(CH_2)_a$) respectively. On addition of salt, both symmetric and asymmetric stretching mode shift their position indicating evidence of the interaction of salt with polymer host.

The deconvolution of the peak is done to get the original peaks in given wavenumber range by Voigt Amp fit. The symmetric and asymmetric stretching mode shifts to lower wavenumber side from 2885 $cm^{-1}$ to 2883 $cm^{-1}$, 2910 $cm^{-1}$ to 2903 $cm^{-1}$ on the addition of nanofiller in PS complex respectively. The addition of nano filler affects the symmetric/asymmetric mode as observed in terms of change in peak area and it may be due to theinteraction of a surface group of nanofiller with polymer host. This shift in band position shows the effect of nanofiller on the interaction between polymer-salt, and polymer-salt-nanofiller in solid polymer electrolytes. FTIR spectrum provides direct evidence of interaction (polymer-ion-nanofiller interaction) in SPEs films. Generally, these interactions impact strongly the dissociation of salt which results in more free ions and results in faster ion dynamics. Also at higher nanofiller content upward shift in wavenumber is shown by asymmetric mode (2914 $cm^{-1}$ → 2910 $cm^{-1}$ → 2909 $cm^{-1}$ → 2903 $cm^{-1}$) which provide evidence of chain stiffening. This is further studied in the following section and is correlated with electrical properties.

### 3.4. Differential scanning calorimetry (DSC) analysis

DSC was implemented for confirming the increase of electrical conductivity which is due to increase of free ions and decrease of crystallinity. Figure 11 (I & II) shows the DSC results for solid polymer electrolytes with different nanofiller content. The endothermic peak observed at high temperature is attributed to the melting point of solid polymer electrolyte. The percentage of crystallinity ($X_c$) was calculated as $X_c = \frac{\Delta H_m}{\Delta H_m^o} \times 100$ where $\Delta H_m$ is the melting enthalpy obtained from the DSC measurement and $\Delta H_m^o$ is the melting enthalpy of pure 100% crystalline PEO (188 J/g) [59]. The melting peak of the polymer blend PEO-PVC observed at 79 °C shift toward lower temperature of 69 °C on the addition of the salt (Figure 11 (I)). Also, the area under the peak decreases which depicts the decrease of crystallinity in the polymer electrolyte.  As polymer crystallinity is hindered due to coordinating interaction of polymer chain with cation. This lowering of the melting temperature and melting peak area is strong evidence of enhanced amorphous content. Also, the enthalpy of melting and melting temperature follows the same trend with the addition of the nanofiller at lower content and this type of trend evidence the increased polymer chain flexibility and ion mobility [60]. The area under the melting curve was used to calculate the degree of crystallinity and glass transition temperature alongwith melting temperature of SPEs are summarized in Table 4. It may be concluded that addition of nanofiller decrease the crystallinity and is due to surface interaction provided by nanofiller with polymer host and the lithium salt.

Now, in figure 11 (II) the slight downward transition in the thermal curve with mid-point in the range -30 to -60 °C reflect the glass transition temperature ($T_g$) of solid polymer electrolytes and shows reduction with the addition of the salt. Further the addition of the nanofiller demonstrates left shift on the x-axis and suggest the increase in conductivity due to the increase of polymer flexibility (Table 4). As nanofiller with high surface area weakens the interaction of cation with the ether group and supports the fast ion migration via the additional conducting pathways provided by the nanofiller surface. The $T_g$ corresponding to PVC is not observed, and it may be due to overlapping of the peak with melting peak of PEO. So, the presence of single glass transition temperature and a shift in glass transition temperature of PEO for polymer blends confirms better miscibility of both polymers [61-62]. The polymer miscibility is due to the presence of strong intermolecular interactions like hydrogen bonding, between the chlorine atoms of PVC with PEO, having non-bonded electrons, like oxygen. The presence of strong interaction reduces the Gibbs free energy which is an indication of homogenous mixing of polymers [63-64]. Next section further correlates the DSC parameters with electrical properties.

### 3.5. Complex Impedance Spectroscopy (CIS) Analysis

Complex Impedance spectroscopy (CIS) is a powerful tool for investigating the electrical properties of materials and motion of bound/mobile ion charges in the solid or liquid materials using a small signal (~ 50 mV) in the frequency domain from 1 Hz to 1 MHz. The principle of the impedance spectroscopy is based on the ability of a medium to pass an alternating electrical current. When an alternating electric field is applied across the sample, dipoles interaction with the corresponding ions due to the Coulomb electric force leads to rearrangement depending on the mobility of backbone ($Li^+$) [53]. The experimental impedance response, when fitted using non-linear least squares (NLS) model by means of a computer program ($Z_{SimpWin}$), agrees well with the corresponding theoretical pattern suggesting validity of the experimental results and an equivalent circuit is proposed to interpret the complex impedance spectra

Figure 12 (a-h) represents the complex impedance plot of prepared SPE films at different nanofiller concentration. For polymer blend semicircle curve is obtained with a diameter $R_b$ extending along the real axis from the origin as in Figure 12 (a). Figure 12 (b) shows typical complex impedance spectra of the blend polymer salt complex. It is usually true that high-frequency response conveys information about properties of an electrolyte such as bulk resistance/long range ordering (due to the migration of ions). The low-frequency response carries information about the electrode/electrolyte interface. The effect of nanofiller addition into the blend polymer salt complex matrix appears to modulate the electrical behavior of composite matrix. This change is designated by the changing pattern of complex impedance spectrum with various nanofiller concentration. The value of bulk resistance decreases with the addition of salt as compared to polymer blend and further addition of nanofiller decrease the bulk resistance. At low frequencies, there is sufficient time to accumulate the space charge across the electrode-electrolyte interface and that results in an enormous increase in the measured double layer capacitance [65-66]. Such modification like a semicircular arc at high-frequency and a low-frequency steep spike in the impedance spectrum of the SPE films suggests a drastic change in the electrochemical properties of the PS film on the addition of nanofiller. The high-frequency semicircular arc at high frequency in the polymer blend is attributed to the contribution due to the presence of resistance and capacitance in the bulk material of sample. The absence of the semi-circular portion (Figure 12 (b-h)) may be attributed due to the fact the corresponding characteristic frequency is higher than the frequency 1 MHz. The presence of steep spike (non-vertical) at an angle less than 90° to the real axis in low-frequency region suggests the formation of space charge at the electrode-electrolyte interface region, commonly referred as the double layer capacitive effect due to non-faradic process occurring at the interface of SPE films and stainless steel electrode [13, 67-69]. It also indicates the inhomogeneous nature of the electrode–electrolyte interface. This type of behavior suggests that migration of ions occurs via free volume available in the polymer matrix and can be represented by a resistor [70-71]. The low-frequency response of CIS pattern remains almost identical for the SPE films as in the case of PS film irrespective of nanofiller variation. It represents the accumulation of space charge carriers at the interface of the ionically conducting SPE films and stainless steel (SS) blocking electrode. The disappearance of the semicircular region in the impedance spectra after addition of nanofiller leads the conclusion that total conductivity is due to conduction of fast ion charge migration, so only diffusion process takes place [72-75]. This also indicates that conduction is now supported by dispersed nanofiller with the surface group.

The experimental impedance response, was fitted using non-linear least squares (NLS) model by means of a computer program ($Z_{Simp}$Win). An electrical equivalent circuit model (ECM) of the sample impedance response appears to be consistent with a fit containing two parallel combinations of constant phase element (CPE) and a resistance connected in series with each other. The presence of a CPE in the material sample evidence of its multiphase character encompassing of microstructure having both crystalline, amorphous anda mixture of the two phases in the PS complex film and $Z_{CPE} = \frac{1}{Q_o(jw)^n}$ , where $Q_o$ and n are fitting parameters and n=0, 0.5, 1, and -1 for pure resistor, Warburg component, pure capacitor, and pure inductive component, respectively. From the fitting pattern of CIS (Figure 12) it is clearly visible that the theoretical and experimental pattern are in good agreement with each other. Change in fitting parameters with filler concentration shows change in electrical properties [76-78].

The parameters recorded in Table 5 are just to provide evidence for a change in the sample electrical response to change in nanofiller concentration in the SPE films. The model of the electrical equivalent circuit remains identical in the pattern for the SPE and PS films with vital difference in the values of bulk resistance ($R_b$), constant phase elements(CPE) $Q_1$ and $Q_2$ and their exponent's $n_1$ and $n_2$, respectively. A high Double layer capacitance ($C_{dl}$) value in the SPE film with nanofiller (53.1 µF) as compared to polymer salt (0.18 µF) film may be attributed to more dissociation of lithium salt due to surface interaction of nanofiller with salt and produce more number of charge carriers for conduction which is in agreement with deconvolution of FTIR.

### *3.5.1. Electrical Conductivity Analysis*

To investigate the effect of nanofiller on polymer blend different content of nanofiller were added to prepare the electrolyte. So, for electrical conductivity measurements using two blocking electrodes, comprising of *SS*|*SPE*|*SS* the equation 1 is used. The ionic conductivity of polymer blend was 3.3 $\times 10^{-8}$ S $cm^{-1}$. The electrical conductivity increases with the addition of salt due to the availability of ions for migration and is a common feature of polymer electrolytes. Initially, at low nanofiller content, there is a small increase in conductivity value but with the addition of nanofiller increase is more due to release of more free ions via interaction of nanofiller-polymer-ion. Further, the addition of nanofiller enhances the salt dissociation and continues increase is seen [79]. The maximum value of conductivity was obtained for 7 wt. % nanofiller (~ $5\times 10^{-5}$ S $cm^{-1}$) which was three orders higher than polymer blend and attributed to the complete dissociation of salt in the system as directed by the FTIR spectra. Also, the maximum conductivity value may be correlated with the structure of the SPEs including the uniform distribution of the nanofiller in the polymer salt matrix as seen by FESEM and EDS mapping. The addition of nanofiller also makes polymer chain flexible which results in fast segmental motion and increase in conductivity is observed. The characteristic conductivity variation with different nanofiller agrees well with the nanofiller dependent changes in the fraction of the free anion in the SPEs system. Further, to support investigated results correlation of various critical parameters of electrical transport was done, and a model is proposed in a later section.

### 3.6. Dielectric Spectroscopy Analysis

#### *3.6.1. Complex permittivity analysis*

Dielectric analysis of solid polymer electrolyte materials is desirable to investigate in detail the transport of ion and are explained in terms of the real and imaginary parts of complex permittivity ($\varepsilon^*$). The dielectric permittivity describes the polarizing ability of material in the presence of an applied external electric field. As permittivity is a function of frequency, so it is a complex quantity shown below:

$$\varepsilon^* = \varepsilon' - j\varepsilon'' \quad (5)$$

The real part of dielectric permittivity ($\varepsilon'$) is proportional to the capacitance and measures the alignment of dipoles, whereas the imaginary part of dielectric permittivity ($\varepsilon''$) is proportional to conductance and represents the energy required to align the dipoles. Here $\varepsilon'$ is related to the stored energy within the medium and $\varepsilon''$ to the dielectric energy loss of energy within the medium. The real ($\varepsilon'$) and imaginary ($\varepsilon''$) parts of the dielectric permittivity are evaluated using the impedance data by Eqn 6 & 7:

$$\varepsilon' = \frac{Z''}{\omega C_o(Z'^2 + Z''^2)} \quad (6)$$

$$\varepsilon'' = \frac{Z'}{\omega C_o(Z'^2 + Z''^2)} \quad (7)$$

Where $C_o$ ($=\varepsilon_r.A/t$) is the vacuum capacitance, $\varepsilon_r$ is the permittivity of free space and ω is the angular frequency.

Figures 13 (a & b) shows the recorded dielectric constant (ε′) and dielectric loss (ε′′) as a function of frequency with different nanofiller. From the plot, it is seen that ε′ decreases with the increase of frequency for all the systems and reach steady state near 10 kHz for all samples. Moreover, both values are higher at low frequency and decrease with the increase in frequency, and that indicates polarization effects due to charge carriers near electrodes and also due to the dipoles which do not begin to follow the field variation at higher frequencies [80-84]. The incorporation of nanofiller increases the dielectric constant since the growth of a number of charge carriers occurs and enhanced ion mobility is achieved. The dielectric constant is higher for 7 wt. % nanofiller content which is in close agreement with conductivity data and supports the present investigation of fast solid state ionic conductor. The low-frequency response is coined by electrode polarization effect which is due to the formation of electric double layer capacitances due to free charge build up at the electrolyte/electrode interface. While the higher frequency region restricts dipoles contributing to electrode polarization due to insufficient response time on the application of electric field and dielectric constant decreases (ε′). Figure 13 (b) shows the tremendous value of dielectric loss (ε′′) towards the low-frequency region and may be attributed to the accumulation of free charge carrier at electrode/electrolyte interface because in this region charge carriers get sufficient time to accumulate at electrode/electrolyte interface and contribute to the large dielectric loss.

*3.6.2. ac conductivity analysis*

The AC electrical measurements (ac conductivity) of PEO-PVC based solid polymer electrolyte were obtained at a frequency range from 1 Hz to 1 MHz at room temperature. The frequency variation of real part of conductivity is shown in Figure 13 (c). The standard feature indicates that ac conductivity increases with increasing the applied frequency due to fast ion migration which may be attributed to increased number of the mobile charge carrier’s and results in charge build or electrode polarization (EP). The frequency dependent real part of conductivity shows three distinct regions, (i) low-frequency dispersion region, (ii) frequency independent plateau region and (iii) high-frequency dispersive region. The low conductivity value at the low-frequency dispersion region is the result of the accumulation of ions (electrode polarization) due to the slow periodic reversal of the electric field. The intermediate region at slightly higher frequency corresponds to the frequency independent plateau region, and dc conductivity can be evaluated from here. At the high-frequency window, the region is due to short range ion transport associated with ac conductivity. Polymer blend and polymer salt system shows all the three regions, but the low-frequency region disappears with the addition of salt and nanofiller. All system shows a shift in both intermediate frequency region and high-frequency region towards high frequency window. The dc conduction is attributed to random hopping of ions between localized states, while the cause of AC conduction is correlated ion hopping in high frequencies region. For all nanofiller, dispersed polymer electrolytes (Figure 13 c), high-frequency dispersion region corresponds to bulk relaxation phenomena and falls outside the measured frequency range hence, could not be observed. This observation is indicative of complexion transport process possibly due to the combined effect of space charge polarization (electrode polarization at the electrode-electrolyte interface) at low frequency followed by long-range ion migration at high frequency [81]. AC conductivity is evaluated by equation 8:

$$\sigma_{ac} = \omega \varepsilon_o \varepsilon'' \quad (8)$$

Where $\omega$ is the angular frequency, $\varepsilon_o$ is the dielectric permittivity of the free space and $\varepsilon''$ represents the dielectric loss.

The conductivity of polymer blend is very low as observed in figure 13 (c) and is due to electrode polarization effect causing a decrease in a number of free charge carriers. The addition of salt shifts the curve toward high-frequency region which is evidenced by an increase in conductivity. The electrode polarization contribution in conductivity is clearly observed by the high dispersive effect which is typical to a fast ion conductor and saturation limit of electrode polarization effect, evidence change in the mechanism of ion migration. The high frequency dispersive region observed in polymer blend and polymer salt system follows universal Jonscher's power law (JPL) response given by $\sigma' = \sigma_{dc} + A\omega^n$, where $\sigma'$ and $\sigma_{dc}$ are the AC and DC conductivities of electrolyte, while A and n are the frequency independent Arrhenius constant and the power law exponent, respectively. Basically, ion migration at low frequency results in long relaxation time, and ion contribution is seen in dc conductivity. Low chi-squared value of order of $10^{-13}$ suggests best JPL fit for the experimental results and fitted values are shown in inset (Figures 13 d & e). The frequency-dependent region fitting provides an estimate of the pre-exponential factor (A) and fractional exponent (n). Value of n between 0.5 and 1 suggests that SPE system is a true ionic conductor. The power law behavior is a universal property of materials.

### 3.7. Transference Number Analysis

The transference number of an ion in SPE is the fraction of the total current carried by the respective ion across a given medium. As, different ions have different mobility's and may bring drastically different portions of the total current [85]. The total active ionic transference number of SPEs has been studied by separation of ion and electronic contribution. The variation of current as a function of time for the PS film and SPE films with varying nanofiller concentration ($x$ wt. % $TiO_2$; $0 \leq x \leq 20$ ) at room temperature is recorded as shown in Figure 14.

The current decays immediately and asymptotically approach steady state. The pattern shows two current one is the initial current, total current ($i_t$) and after that, there is a sharp drop in the value of current with the passage of time. After some time current reaches to saturated value and same trend of constant current is obtained for all the samples. The two processes of migration of ions under the influence of an external field and diffusion due to concentration gradients are hostile, and therefore after sufficiently long time, the establishment of steady state is reached [17]. Initially in SPE membrane high current is attributed to migration of both electrons and ions while current after cell polarization due to blocking electrodes (SS) is attributed to movement of electrons only, known as the residual electronic current ($i_e$). The initial decrease in current until the saturation state is primarily due to the formation of the passivation layer on electrodes. Now concentration gradient of ions develops and opposes the applied current by diffusion of ions. As the polarization builds up because of the applied voltage, the ions were blocked at the blocking electrode there by preventing the ionic current ($i_{ion}$), and the last current comprises only the electronic current ($i_e$) [86]. The high value of transference number also indicates a reduction in concentration polarization. This improves the ionic transport, and after a long time, the presence of the steady state evidences the current flow only due to the migration of $Li^+$ cations in the electrolyte and anion movement has been entirely ceased [87-88]. The estimated value of the ionic transport number ($t_{ion}$) using experimental data of $i_t$ and $i_e$ is obatined with the eq. (2) and given in Table 6. Further ionic and electronic conductivity contribution is estimated using equation 9-11.

$$i_t = i_{ion} + i_{elec} \qquad (9)$$

Therefore

$$\sigma_{ionic} = \sigma_{electrical} \times t_{ion} \qquad (10)$$

$$\sigma_{electronic} = \sigma_{electrical} \times t_{electronic} \qquad (11)$$

The transference number values clearly indicate that the SPE samples are a chiefly ionic conductor with $t_{ion} \approx 0.99$ and negligible electronic contribution and is sufficient to fulfill the requirement of solid state electrochemical cells [89-90]. The ionic conductivity and electronic conductivity are also in absolute correlation with electrical conductivity value. Beside this, the electronic conductivity is almost within the desired limit ($\leq 10^{-7}$ S cm$^{-1}$) for its utility in energy storage devices.

### 3.8. Electrochemical Stability Window

A wide electrochemical stability window of the electrolyte is an important parameter for the effective performance of energy storage/conversion devices. The working voltage or electrochemically stability window of the SPEs membranes has been studied by observing the variation of current and voltage (I & V) using LSV technique (Figure 15) for PS films and SPE films with nanofiller. Initially, there is almost constant current through the electrodes up to 2 V, while after that with an increase of the voltage there is a slow increase in current followed by an abrupt change corresponding to the onset of the decomposition process of the electrolyte. This small current upto 2 V might be attributed to the change of the stainless steel surface. This kind of behavior is different from both PS and SPE films and provides direct evidence of enhancement in working voltage or electrochemically stability window after addition of nanofiller as compared to PS films. The voltage value is obtained by extrapolating the higher voltage linear current with the x-axis. During the LSV, the working electrode is polarized, and the onset of the current may be taken as the decomposition voltage of the given polymer electrolyte [91]. The addition of nanofiller in PS films gives a safe operating range with the majority of the lithium battery electrode couples [7]. A comparison of this with PS films indicates an enhancement in voltage window in SPE films (2.8V to ~3.5 V) with a maximum value of ~3.5 V. Of course, the voltage stability window enhancement appears to be a function of wt. % of nanofiller evident from the results.

### 3.9. Thermo-gravimetric analysis

Now, the thermal stability of the prepared polymer nanocomposite films is investigated by TGA. As during operation of a battery device especially in electric vehicles, high temperature may reach which may decompose the constituents of the polymer electrolyte and may lead to explosion [92]. So, it becomes the principal responsibility to investigate the thermal stability of the materials used in the system. Figure 16 represents the thermo-grams of PEO-PVC blend and PEO-PVC+$LiPF_6$ with 0 wt. % $TiO_2$, 7 wt. % $TiO_2$ and 15 wt. % $TiO_2$. For better in sights thermo-gram is divided into three regions depending on the weight loss.

In the *region 1* initially, a small weight loss is observed and that may be due to the evaporation of the solvent present in the skeleton of the films. PEO-PVC blend with salt shows a reduction in thermal stability as compared to the PEO-PVC blend and it may be due to more aprotic solvent absorbed by the salt. Also the coordinating interaction between the cation and ether group affects the thermal stability of polymer salt complex. While, the addition of nanofiller indicates a reduction of solvent content and better thermal stability is observed as compared to the polymer salt system. As dispersion of nanofiller alters the interaction between polymer chain and a cation that leads to the creation of some new interaction between the nanofiller and cation which affects the thermal stability. Now, in *region 2* PEO-PVC blend shows rapid weight loss (apprx. 70 %), while the addition of salt prevents the rapid loss and thermal stability is improved. This may be attributed to the cation coordination with the ether group of the polymer chains. This may be due to the breakage of weak interaction and materials starts decomposing. While, for polymer salt system with nanofiller, thermal stability curve shifts right side or thermal stability is improved. All films are and thermally stable up to apprx. 300 °C and is considerable for application purpose. This multi-step process of weight loss indicates that the present system is blend polymer electrolyte. In the *region 3*, the polymer matrix constituents start to degrade and weight loss is maximum. But, the solid polymer electrolyte cannot be completely decomposed even when the temperature reaches to 600 °C. All films are thermally stable up to apprx. 300 °C and is enough to fulfill the demand of electrolyte in energy storage/conversion devices.

### 3.10. Correlation of free ions area (%), $C_{dl}$, $\sigma_{dc}$, n, μ, D, $T_g$ (°C), $T_m$ (°C), $\Delta H_m$ (J/g) and $X_c$ (%)

The electrical conductivity value of solid polymer electrolyte films is directly linked with number of free charge carriers and mobility as in given equation 12:

$$\sigma = nq\mu \qquad (12)$$

Where n is a fraction of free ion, q is an anionic charge, and μ is mobility of ions. As q and μ are constant for a particular matrix, so conductivity only depends on the number of free charge carriers as studied in followed section [81]. An excellent correlation of maxima in a fraction of free ion, double layer capacitance, and conductivity with nanofiller content is in support of our scheme (Figure 17 a-c). The increase in ionic conductivity may be attributed to the existence of an amorphous layer with nanofiller on a surface having a surface group which provides channels for ion transport. The electrical conductivity calculated for different nanofiller content is plotted in Figure 17 d. Double layer capacitance is calculated using equation 13 and is shown in Table 5. As nanofiller concentration increases, a number of free charge carriers are available for conduction, and this leads to increase in capacitance value.

$$C_{dl} = -\frac{1}{\omega z''} \qquad \ldots\ldots\ldots\ldots (13)$$

A comparison of conductivity value suggests an enhancement in conductivity after addition of small content of nanofiller 1 wt. % in polymer salt matrix. This may be attributed to release of more free charge carriers due to polymer-ion-nanofiller interaction. The fraction of the free anions increases initially with the addition of even a small quantity of nanofiller and shows a maximum of ~7 wt. %, followed by a downward trend. Another maximum appears at~15 wt. % of nanofiller particles, followed by a decreasing trend. This two maxima trend can be correlated with the fraction of free ions, double layer capacitance, the number density of charge carriers, mobility and diffusion coefficients observed in FTIR and impedance study discussed above. It may be noted that the variation of the fraction of ion pairs as a function of nanofiller concentration shows a minimum corresponding to each maximum in the free ions vs. nanofiller content plots. Such variations of free and paired anions have been reported earlier in PMMA based solvent-free polymer electrolyte composites [93].

The desirable property of a solid polymer electrolyte is high ionic conductivity (σ) which is directly linked with number density (n), mobility ($\mu$) of charge carriers and diffusion coefficient (D) for any plastic separator stands for electrolyte cum separator. As from conductivity analysis it is concluded that conductivity increases with addition of salt in polymer blend and further addition of nanofiller leads to enhancement with two maxima one at 7 wt. % followed by 15 wt. % and is due to the increase of mobility or number density of charge carriers. The mobility (($\mu$) of ions indicates the degree of ease with which ions pass through media on application of external electrical field and the diffusivity (D) represents the ease with which ions pass through media under a concentration gradient. Both parameters depend on the number of free charges as ion pair formation or association may reduce the above number and mobility. Impedance and FTIR study provides direct calculation of number density of mobile ions and their mobility for performance and development of a good solid polymer electrolyte. Here the above parameters are calculated using the FTIR spectroscopy as proposed by Arof et al., [94]. The variation in conductivity can be related to the number density (n), mobility ($\mu$) and diffusion coefficient (D) of charge carriers in the electrolyte. FTIR deconvolution was done to determine the percentage area of free ion and ion pair and the areas are plotted as a function of various nanofiller (Figure 8). The free ion area increases with addition of nanofiller and first maxima at 7 wt. % followed by another at 15 wt. % as shown in Figure 17 c. The number density (n), mobility ($\mu$) and diffusion coefficient (D) of the mobile ions was calculated using following eqn (14)-(16):

$$n = \frac{M \times N_A}{V_{Total}} \times \text{free ion area (\%)} \quad (14)$$

$$\mu = \frac{\sigma}{ne} \quad (15)$$

$$D = \frac{\mu k_B T}{e} \quad (16)$$

In Eqn (14), M is the number of moles of salt used in each electrolyte, $N_A$ is Avogadro's number ($6.02\times10^{23}$ $mol^{-1}$), $V_{Total}$ is the total volume of the solid polymer electrolyte, and $\sigma$ is dc conductivity. In eqn (15), e is the electric charge ($1.602\times10^{-19}$ C), $k_B$ is the Boltzmann constant ($1.38\times 10^{-23}$ J $K^{-1}$) and T is the absolute temperature in eqn 16. Table 7 lists the values of $V_{Total}$, free ions (%), n, $\mu$, D obtained using the FTIR method [94-95]. It is observed that the ionic conductivity is closely related to the number density of mobile charge carriers, double layer capacitance, mobility, and diffusion coefficient (Figure 17 a, e, j). Figure 17 reveals the one-to-one correspondence between free ion area (%), ionic conductivity (σ), charge carrier number density (n), charge carrier mobility ($\mu$) and diffusion coefficient (D). The thermal parameters $T_g$, $T_m$ obtained from DSC analysis indicates that addition of nanofiller suppresses the polymer chain recrystallization tendency which directly confirms the increase of free volume or amorphous phase content at a lower temperature than polymer blend $T_g$ and $T_m$ (Figure 17 h-i). At very high nanofiller content crystallinity shows a slight increase which is due to the negligible effect of nanofiller on polymer chain and may be attributed to the mutual interaction of nanofiller particles which dominates above the polymer-ion-nanofiller interaction. This result is also in good agreement with the associated FTIR ion-ion interaction and conductivity studies. The trend in variation in free ions area (%), conductivity, double layer capacitance, n, $\mu$ and D, is almost identical. Above results are strongly in favor of desirable SPEs for energy storage devices with the evidence provided by the FTIR and impedance analysis.

### 3.11. Self-Proposed Ion Transport Mechanism

The results and their analysis described above enable us to propose a model to explain the strong dependence of the ion conductivity of the PNC films on nanofiller concentration. On the basis of above data obtained by FTIR spectroscopy on polymer–ion–nanofiller interaction and excellent correlation with double layer capacitance ($C_{dl}$), conductivity, number density (n), mobility (μ) and diffusion coefficient (D) have provoked us to analyze and propose a model which can provide better insight toward the improvement of conductivity in the present study. Generally, when the lithium salt is added to polymer electrolyte system, then anion stay bound with the polymer chain while cation gets coordinated with ether group of the host polymer. This electron rich group provides a path for ion migration in the polymer matrix. When nanofiller is added in the polymer salt system, then there are two means by which it can affect the ionic transport or conductivity: (i) direct interaction of cation with nanofiller surface, (ii) interaction of nanofiller surface with a polymer chain. These above said ways leads to various possibilities for interaction: (i) cation coordination with ether group (Fig 18 a), (ii) direct interaction of anion with nanofiller via hydrogen bond (Fig 18 c), (iii) weak interactions of basic site of nanofiller with methylene group of polymer (Fig 18 d), (iv) acidic site interaction of nanofiller with ether group via hydrogen bonding (Fig 18 e), (v) strong interactions of basic site of nanofiller with cation (Fig 18 f).

Accordingly, in the present study, the first maxima observed at low nanofiller content is possibly due to the dissociation of free ions by nanofiller interaction with an anion via hydrogen bonding (Figure 19 a). The interaction between anion and nanofiller dominates at lower nanofiller content and anions can be held on the nanofiller surface, and cation easily migrates via the path provided by polymer chain. The first maxima may also, be linked to the formation of additional space charge region between polymer and nanofiller interface in solid polymer electrolyte due to acidic nature of $TiO_2$ nanofiller. Initially, the affinity for anions is more toward nanofiller surface than the cation.

The acidic nature of nanofiller obstructs ion pairing due to the interaction of $TiO_2$ with $PF_6^-$ ($TiO_2+PF_6^- \leftrightarrow TiO_2:PF_6^-$) and a local electric field is formed which assist in salt dissociation. As overall dielectric constant of the polymer matrix is also higher for 7 wt. % nanofiller content and is most efficient for conduction of lithium ions discussed above in dielectric analysis (Figure 13 a). Further, the double layer capacitance also followed the same trend as of conductivity maxima for 7 wt. % nanofiller. This enhancement in conductivity also enhances the mobility and diffusion coefficient of cation in SPE system as shown in Figure 17 e & j. The decreasing trend in conductivity pattern for high nanofiller content is due to the increased ability of ion pairing, as nanofiller also interacts weakly with ether group, hence salt is not dissociated properly. This reduces the free ion charge carrier concentration, and conductivity enhancement is prohibited or lowering of ion mobility (Figure 17 j). The reduction in conductivity value after first maxima may be due to the formation of ion pairs (Figure 19 d) or reduction of double layer capacitance as evidenced by FTIR and impedance study (Figure 17 b & c). The second conductivity maxima observed may be due to the fromation of the conductive path by the nanofiller grains which promotes migration of cation. As -OH group on the nanofiller surface may play an effective role and interact with the anion via hydrogen bonding. It prevents the ion pairing, and free ions are produced accompanied by an increase in conductivity or the ion mobility (Figure 19 c & d).

At high nanofiller content decrease in conductivity after second maxima may be due to two causes: (i) nanofiller aggregation, (ii) reduction of the suitable coordinating site as represented in Figure 19 e & f respectively. Former one leads to localization of cation and anion in the region of the cluster which reduces the free ion for migration. Hence conductivity decreases as discussed in FTIR spectra (Figure 17 b). The entrapment of ion in localized clusters may be due to high surface area of nanofiller and space charge layer leads to blocking effect. Also,the presence of mutual interaction between nanofiller surface may trap the cation and confine it in between the nanofiller aggregation region, hence further forward movement of the cation is prevented. The immobile state achieved by the polymer chain decreases the conductivity due to termination of conducting pathways for the ion migration and is in agreement with the mobility and free ion area (Figure 17 j and Figure 17 c). This does not support ion conduction. Also at high nanofiller content, one possible reason may be a reduction of coordinating site for lithium may which is due to the interaction of $Ti^{4+}$ with ether group of PEO, hence drop in conductivity is evidenced (Figure 19 f).

## 4. Conclusions

Free-standing solid polymer electrolyte films based on (PEO-PVC)-$LiPF_6$+ *x* wt. % $TiO_2$ have been prepared by solution cast technique. The XRD results reveal the occurrence of complexation between polymer blend, salt and nanofiller. FESEM micrograph provides evidence of change in surface morphology and homogeneity with the addition of nanofiller; further EDS mapping confirms the uniform dispersion of nanofiller in polymer salt matrix. The FTIR analysis confirmed the existence of a polymer-ion, ion-ion, and polymer-ion-nanofiller interactions that improve the electrical transport properties. The FTIR results also indicate an increase in the fraction of free ions with the addition of nanofiller and are in good agreement with the plot of conductivity, double layer capacitance, mobility, and diffusion coefficient. The observed electrical conductivity is highest for 7 wt. % about ~$5\times10^{-5}$ S $cm^{-1}$ at room temperature and appears to be consistent with deconvolution of FTIR results. $TiO_2$ nanofiller contribute to the conductivity enhancement in SPE through the formation of Lewis acid–base type transient bonding and provide additional sites for $Li^+$ ion hopping. The variation in dielectric constant and ac conductivity provides an analysis of the dielectric property, and ac conductivity variation obeys the Jonscher power law. All SPEs exhibit good thermal stability up to 300 °C. DSC analysis shows a reduction in crystallinity with the addition of nanofiller and decrease in glass transition temperature confirms faster segmental motion of polymer chain or enhanced flexibility. The enhancement in ionic transference number and electrochemical potential window on the addition of nanofiller has been estimated. This improvement in electrical conductivity, potential window, ion transference number and mobility of SPE films suggests a prospective candidate for energy storage devices such as lithium polymer batteries and solid state polymeric supercapacitor. A self-proposed ion transport model for enhancement in conductivity has been reported to understand the role of nanofiller and is consistent with experimental data.

**Acknowledgement**

One of the authors acknowledges with thanks for financial support from CUPB and partial funding from UGC Startup Grant (GP-41).

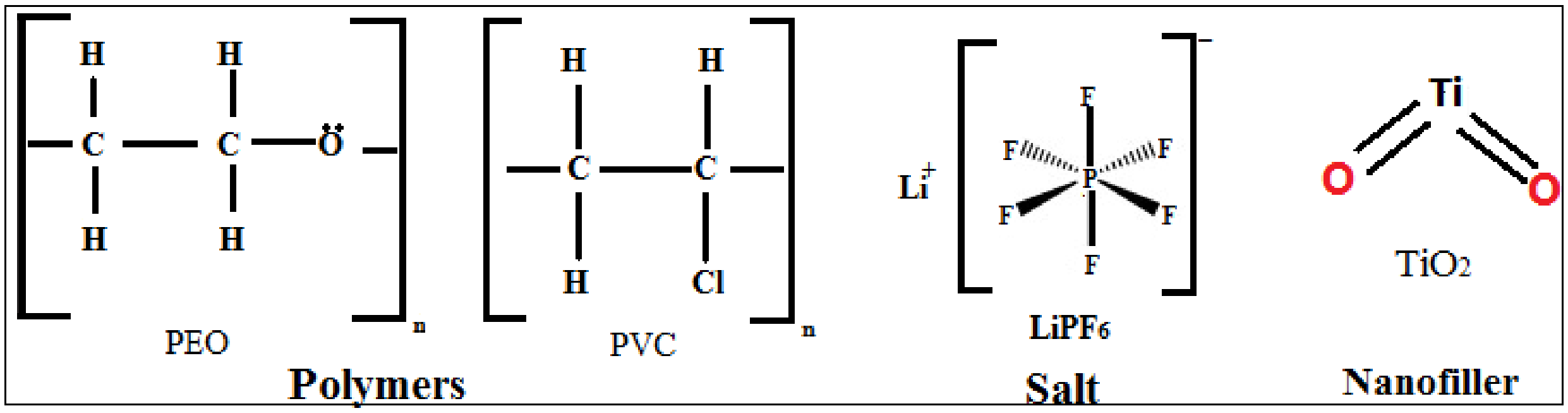

**Figure 1.** Molecular structure of materials

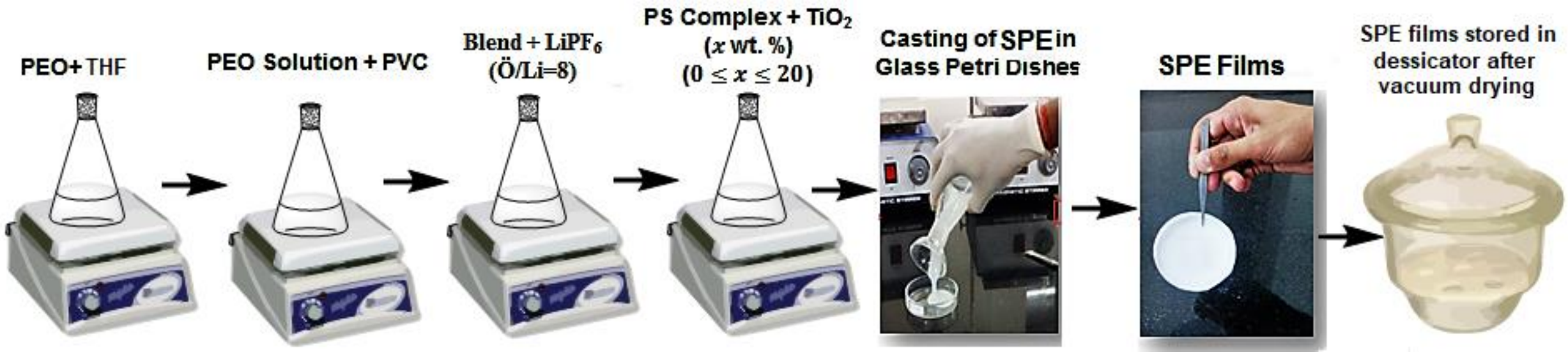

**Figure 2.** Flow chart of solution cast technique

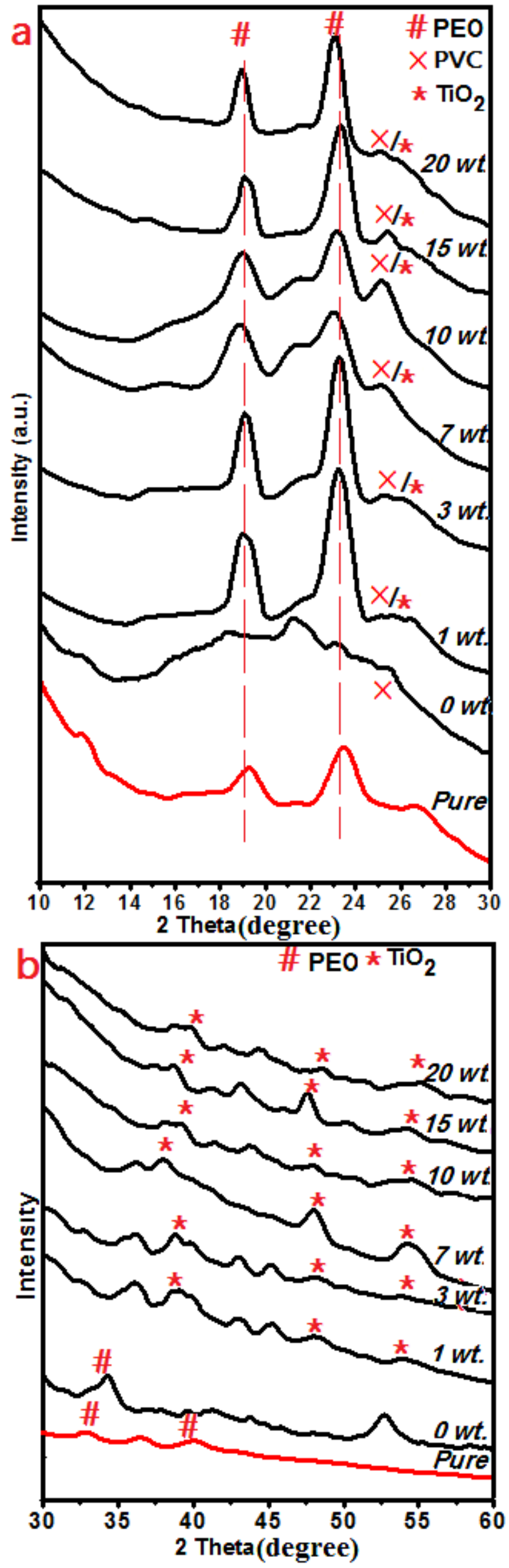

**Figure 3.** XRD patterns of various solid polymer electrolytes with nanofiller $TiO_2$.

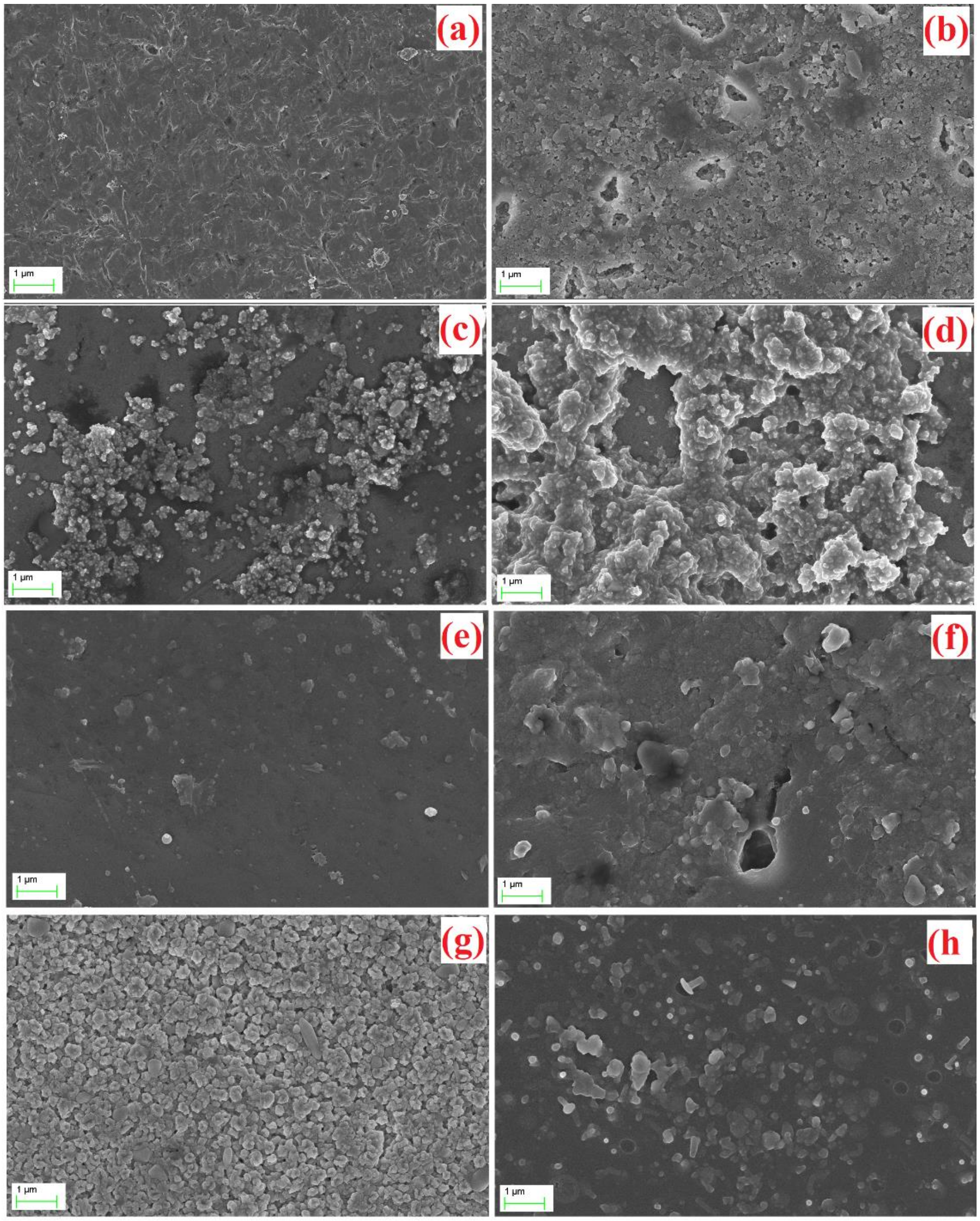

**Figure 4.** FESEM micrograph of SPE films comprising of **(a)** host blend polymer and (PEO-PVC)-$LIPF_6$+ *x* wt. % $TiO_2$ nanofiller **(b)** x =0 **(c)** x=1 **(d)** x=3 **(e)** x=7 **(f)** 10 **(g)** x=15 **(h)** x=20.

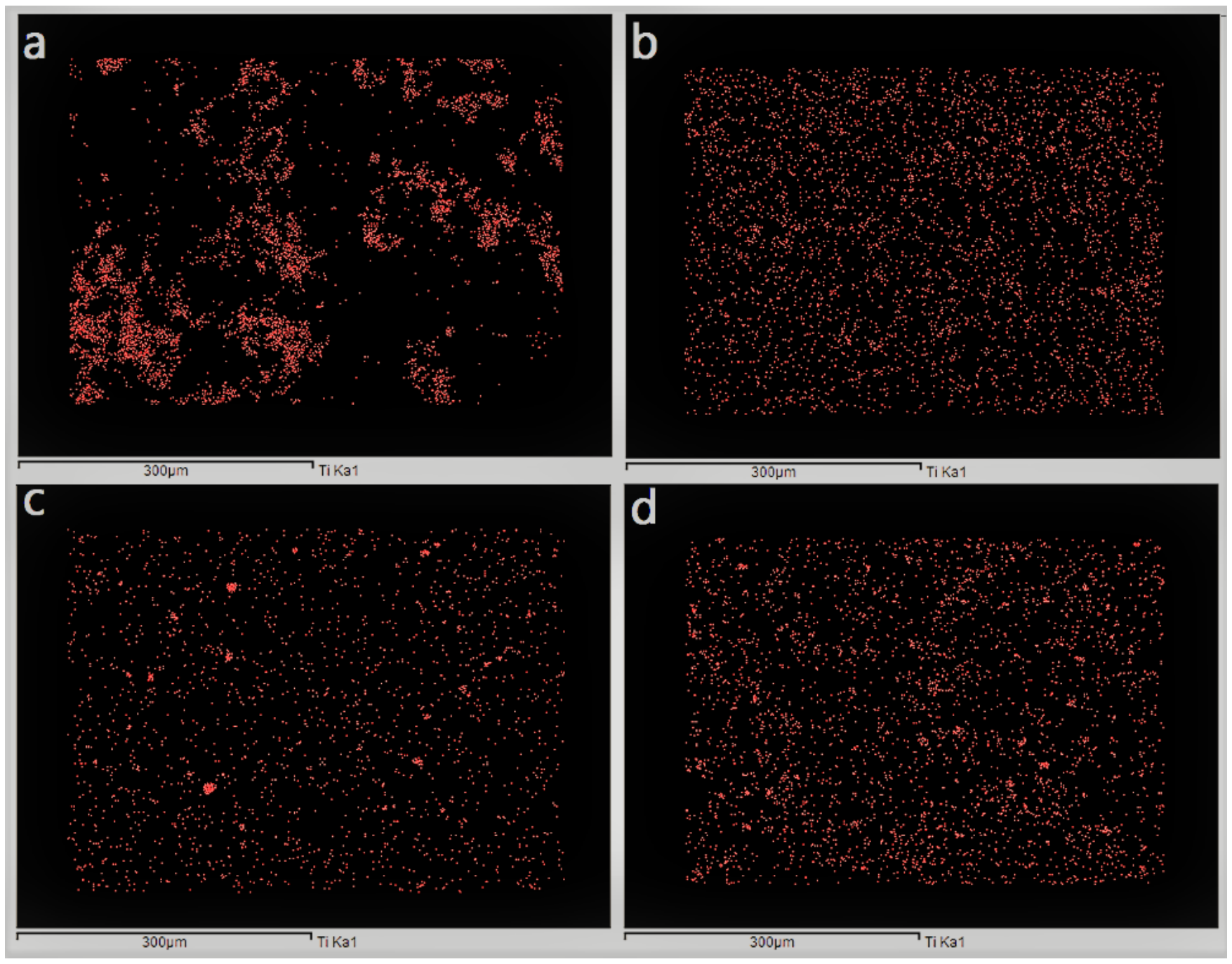

**Figure 5.** Elemental mapping results of Ti atoms in SPEs for *x* wt. % nanofiller (a) x= 3, (b) x= 7, (c) x= 10, (d) x= 15)

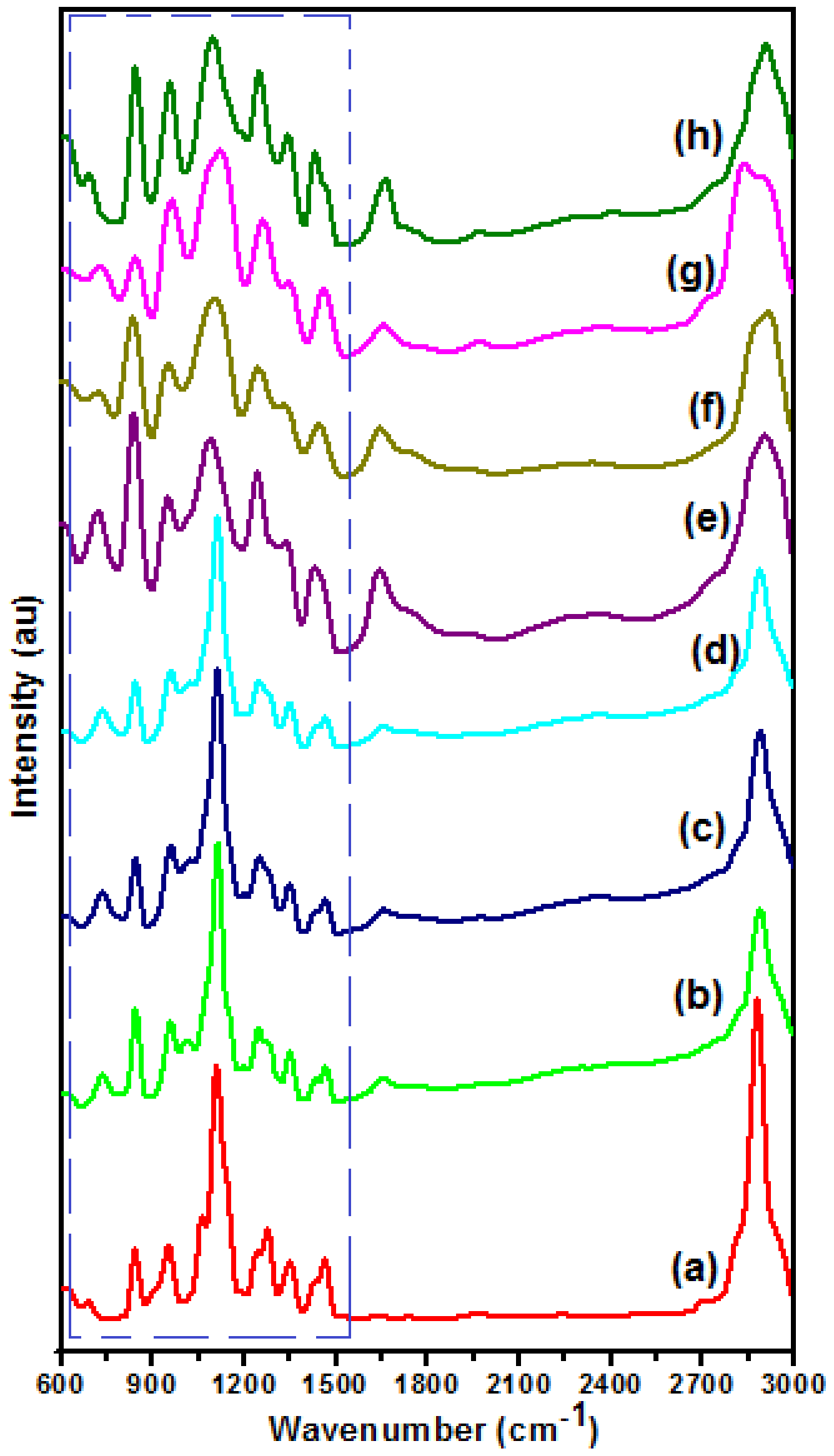

**Figure 6.** FTIR spectra of **a)** blend polymer and (PEO-PVC)-LIPF$_6$+ *x* wt. % $TiO_2$ films **(b)** x =0 **(c)** x=1**(d)** x=3 **(e)** x=7 **(f)** 10 **(g)** x=15 **(h)** x=20

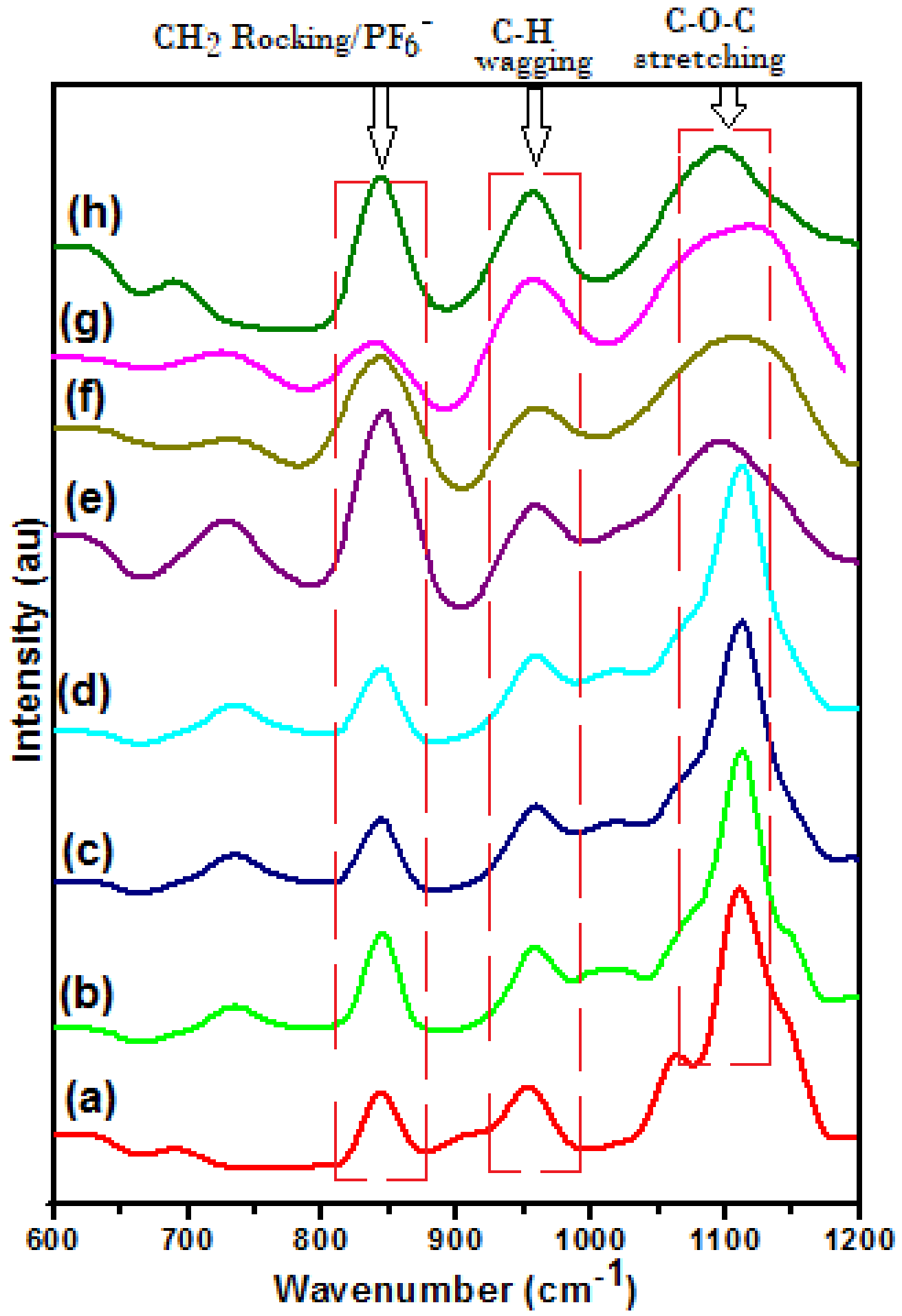

**Figure 7.** Variation in C-O-C peak of PEO in SPE films (a) host blend polymer and (PEO-PVC)-$LiPF_6$+ x wt. % $TiO_2$ films (b) x =0 (c) x=1(d) x=3 (e) x=7 (f) 10 (g) x=15 (h) x=20

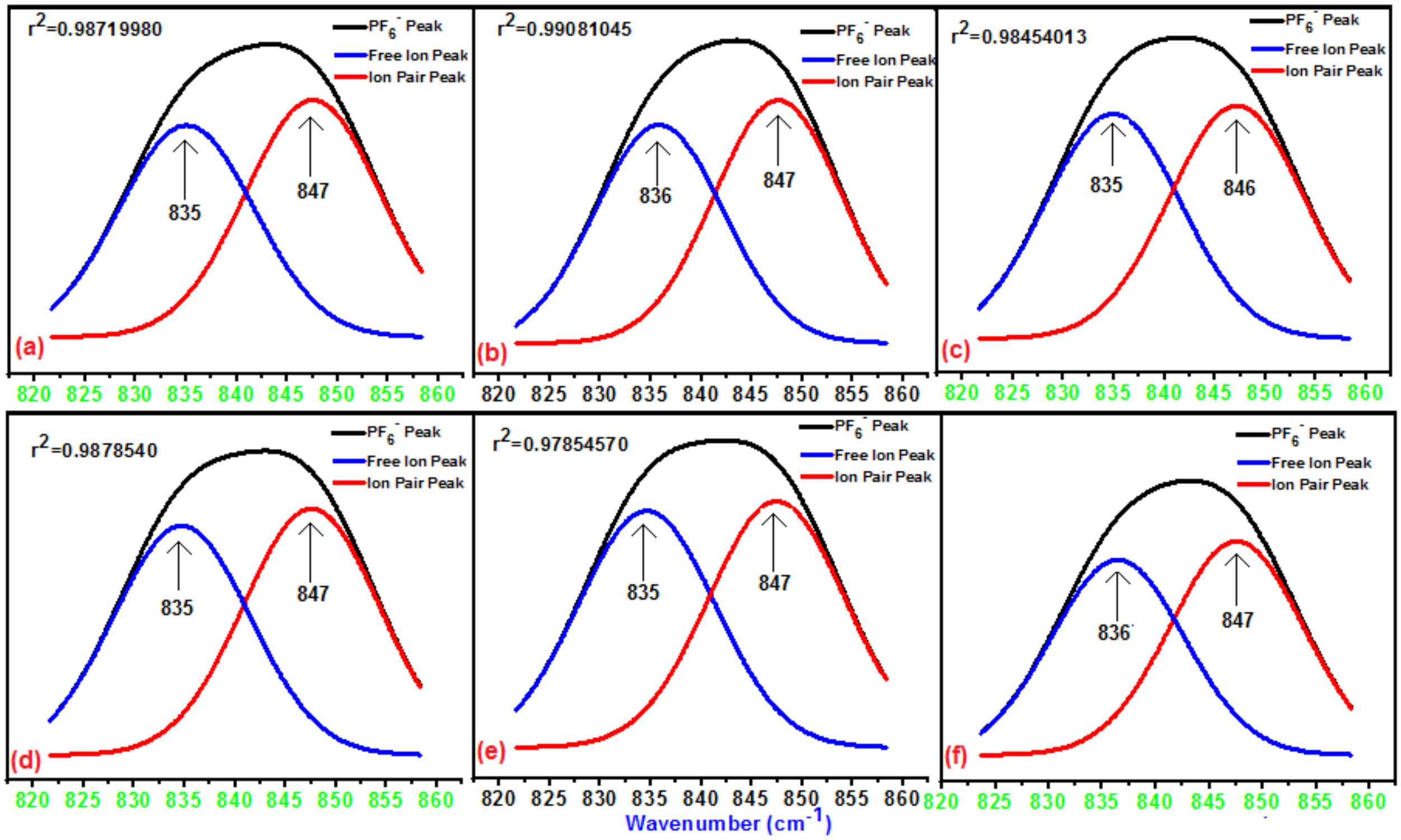

**Figure 8.** Deconvoluted pattern of hexafluorophosphate ($PF_6^-$) band at $\vartheta$=840 $cm^{-1}$ showing contribution of free ion and ion pair in (PEO-PVC)-$LIPF_6$+ x wt. % $TiO_2$ films **(a)** x =0 **(b)** x=3**(c)** x=7**(d)** x=10**(e)** x=15 **(g)** x=20.

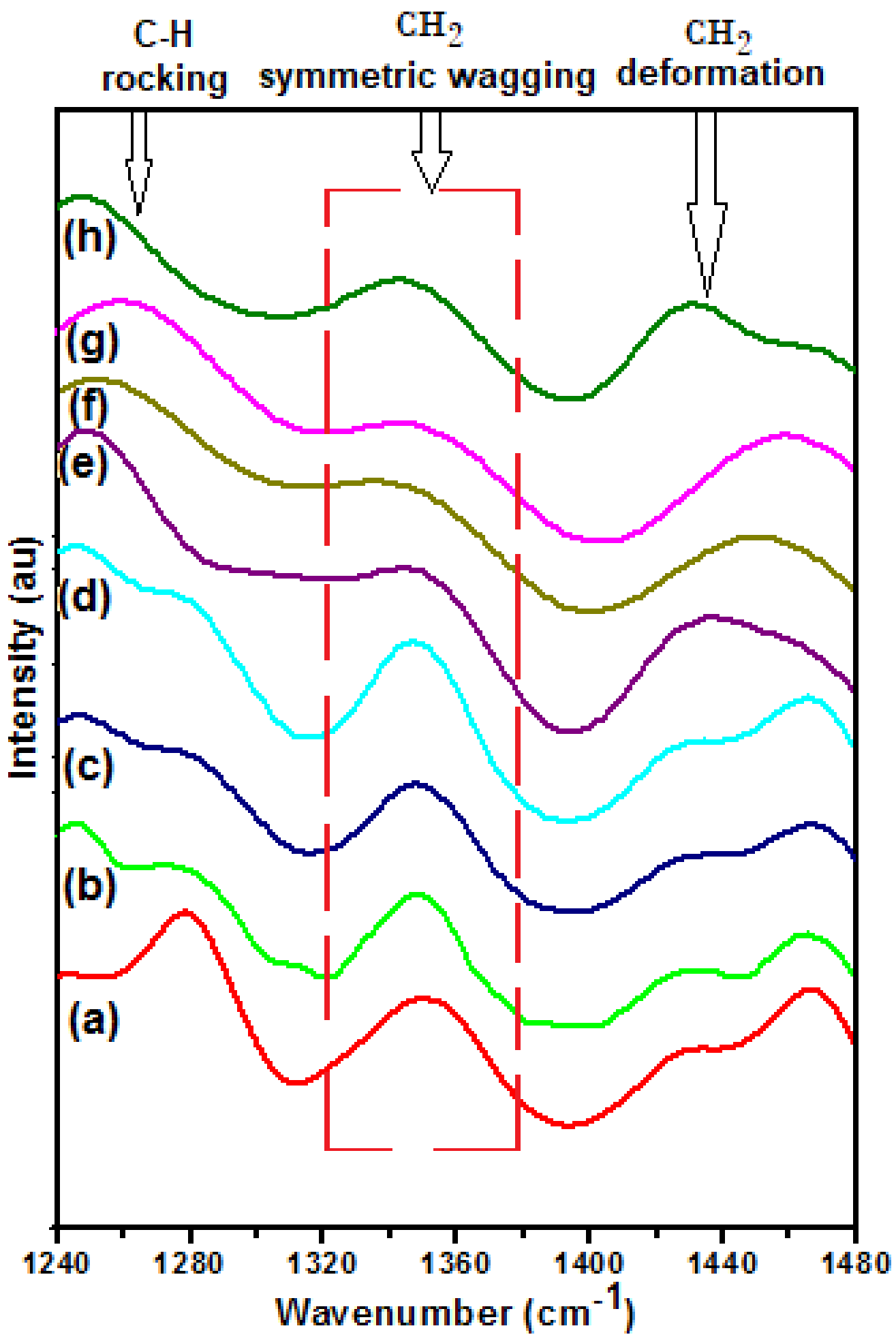

**Figure 9.** FTIR spectra of SPE films **(a)** blend polymer and (PEO-PVC)-$LIPF_6$+ x wt. % $TiO_2$ films **(b)**x=0 **(c)** x=1 **(d)** x=3 **(e)** x=7 **(f)** 10 **(g)** x=15 **(h)** x=20.

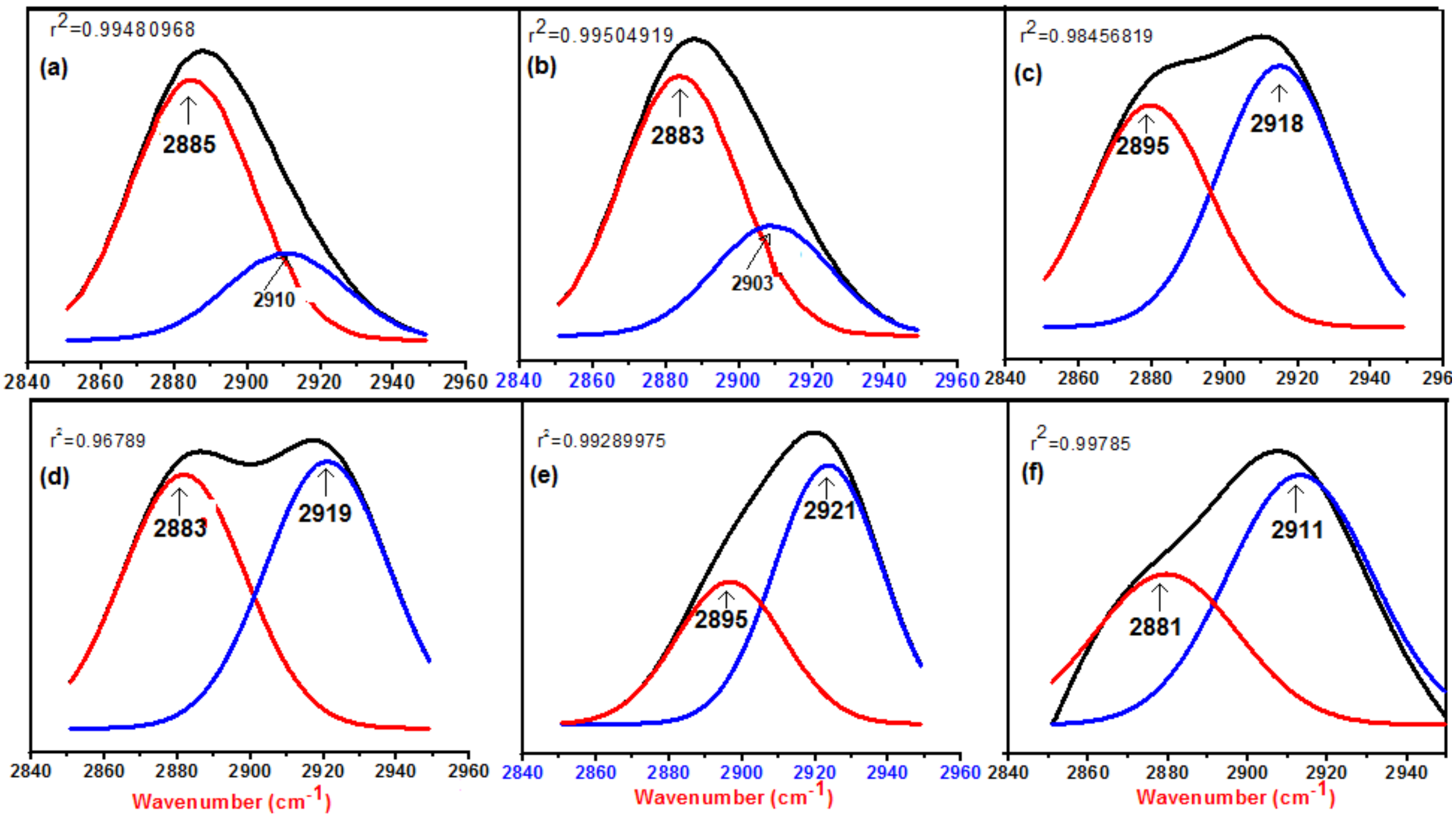

**Figure 10.** Variation in symmetric and asymmetric stretching mode (PEO-PVC)-LIPF$_6$+ x wt. % TiO$_2$ films **(a)** x =0 **(b)** x=3 **(c)** x=7 **(d)** x=10 **(e)** x=15 **(f)** x=20.

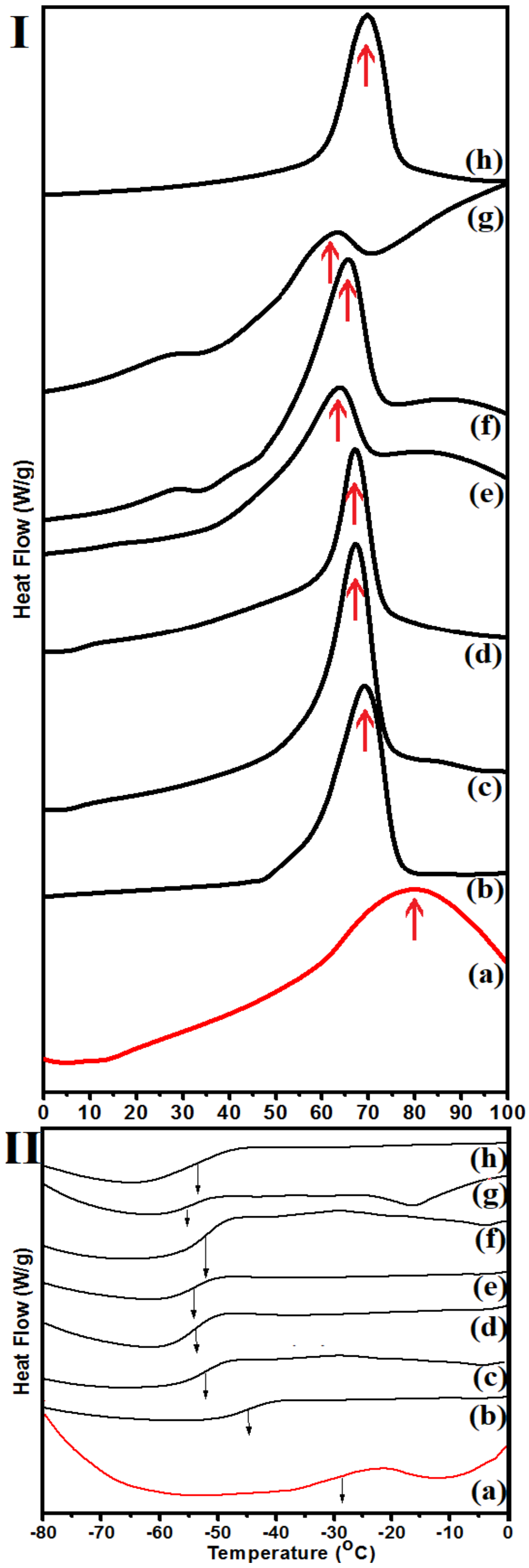

**Figure 11.** DSC curve (I) melting peak, (II) glass transition temperature for various solid polymer electrolytes for films (a) host blend polymer and (PEO-PVC)-$LIPF_6$+ x wt. % $TiO_2$ films (b) x =0 (c) x=1(d) x=3 (e) x=7 (f) 10 (g) x=15 (h) x=20

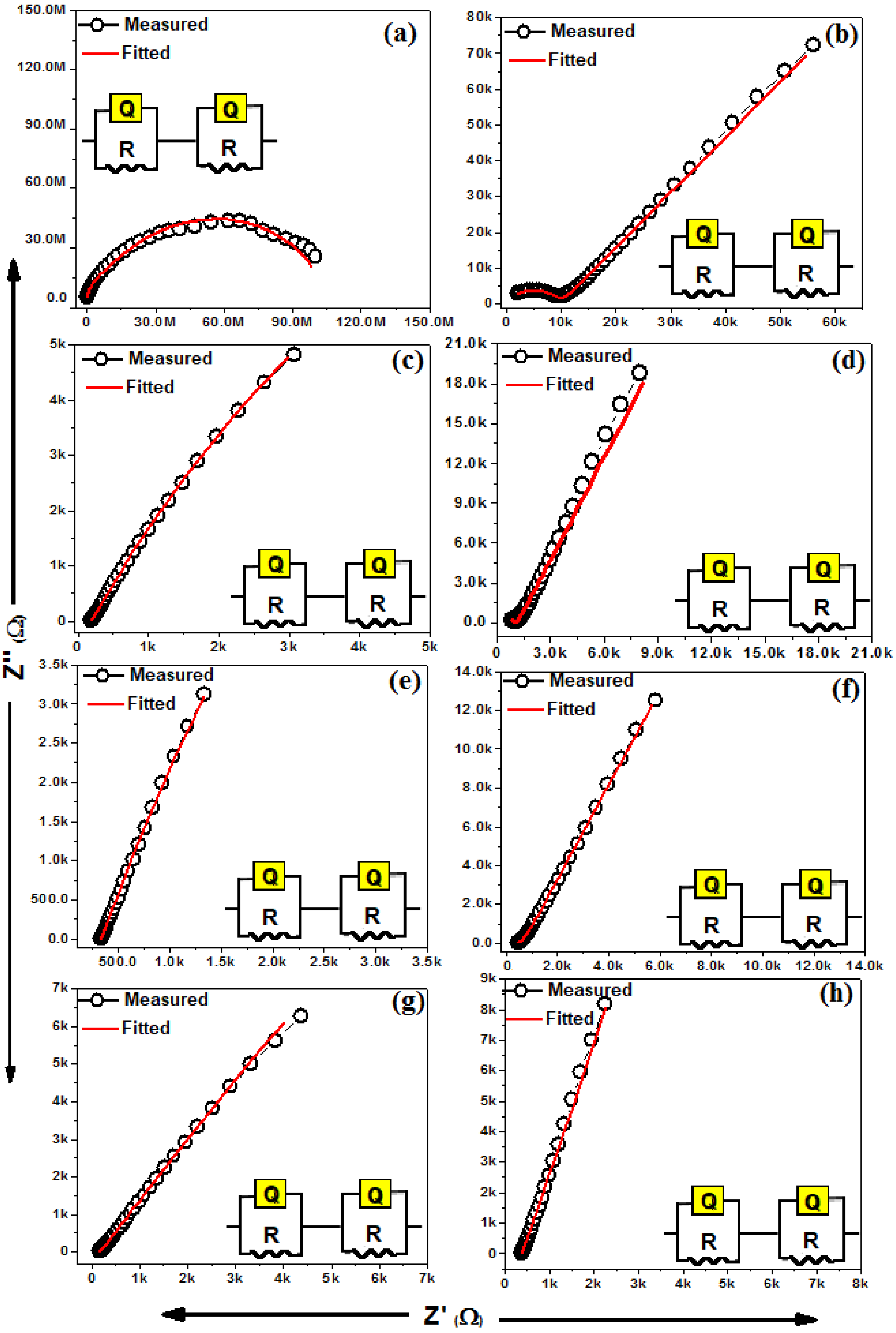

**Figure 12.** Nyquist plot of SPE films **(a)** host blend polymer and (PEO-PVC)-$LIPF_6$+ x wt. % $TiO_2$ films **(b)**x =0 **(c)** x=1 **(d)** x=3 **(e)** x=7 **(f)** 10 **(g)** x=15 **(h)** x=20.

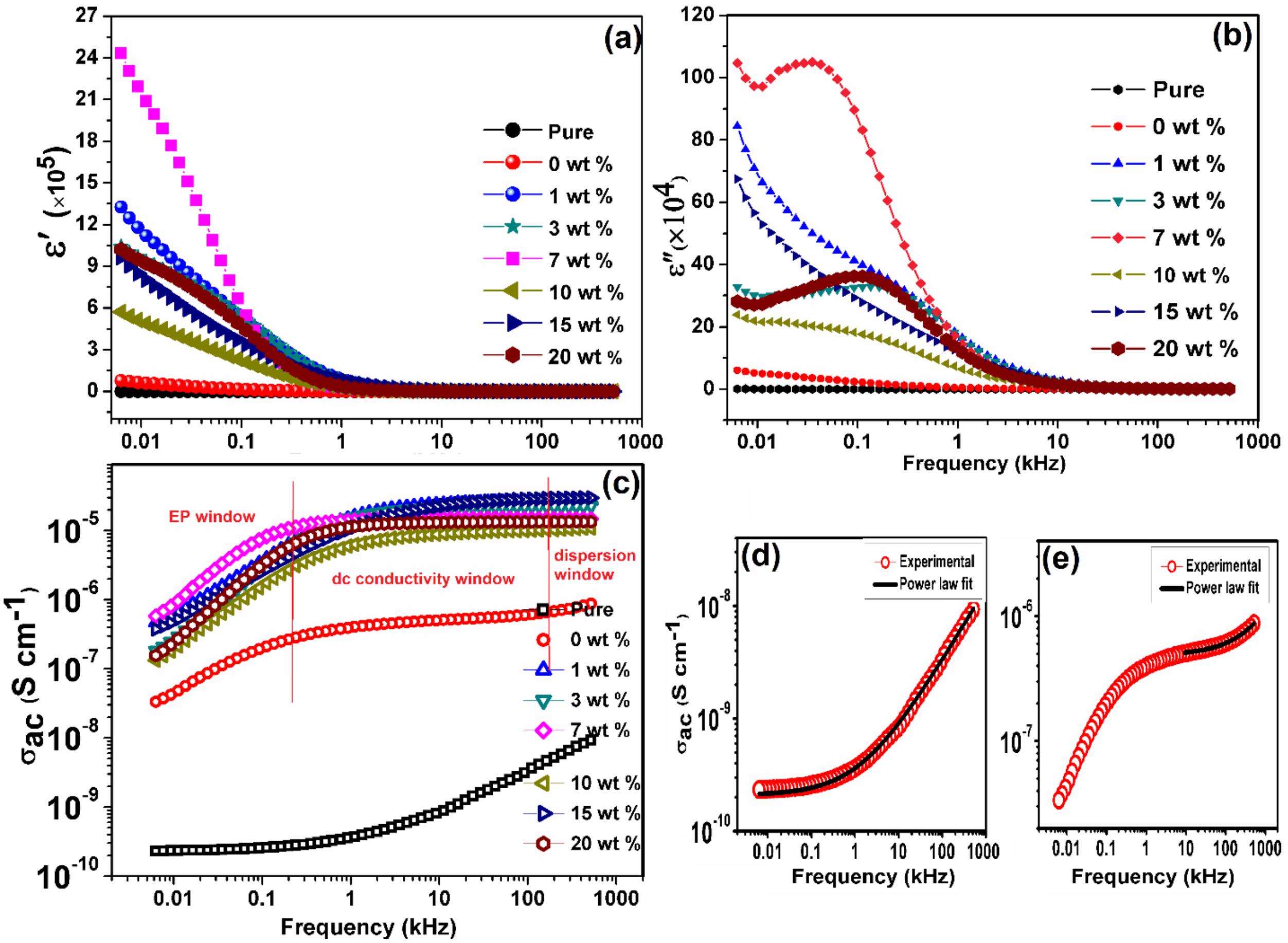

**Figure 13.** Variation of (a) real (ε′), (b) imaginary part (ε″) of dielectric constant, (c) A.C. conductivity against the frequency for (a) PEO-PVC, PEO-PVC+LIPF$_6$- *x* wt. % $TiO_2$ films **(b)**x =0 **(c)** x=1 **(d)** x=3 **(e)** x=7 **(f)** 10 **(g)** x=15 **(h)** x=20 and (d-e) Jonscher power law fit of ac conductivity at high frequency for polymer blend and polymer salt system.

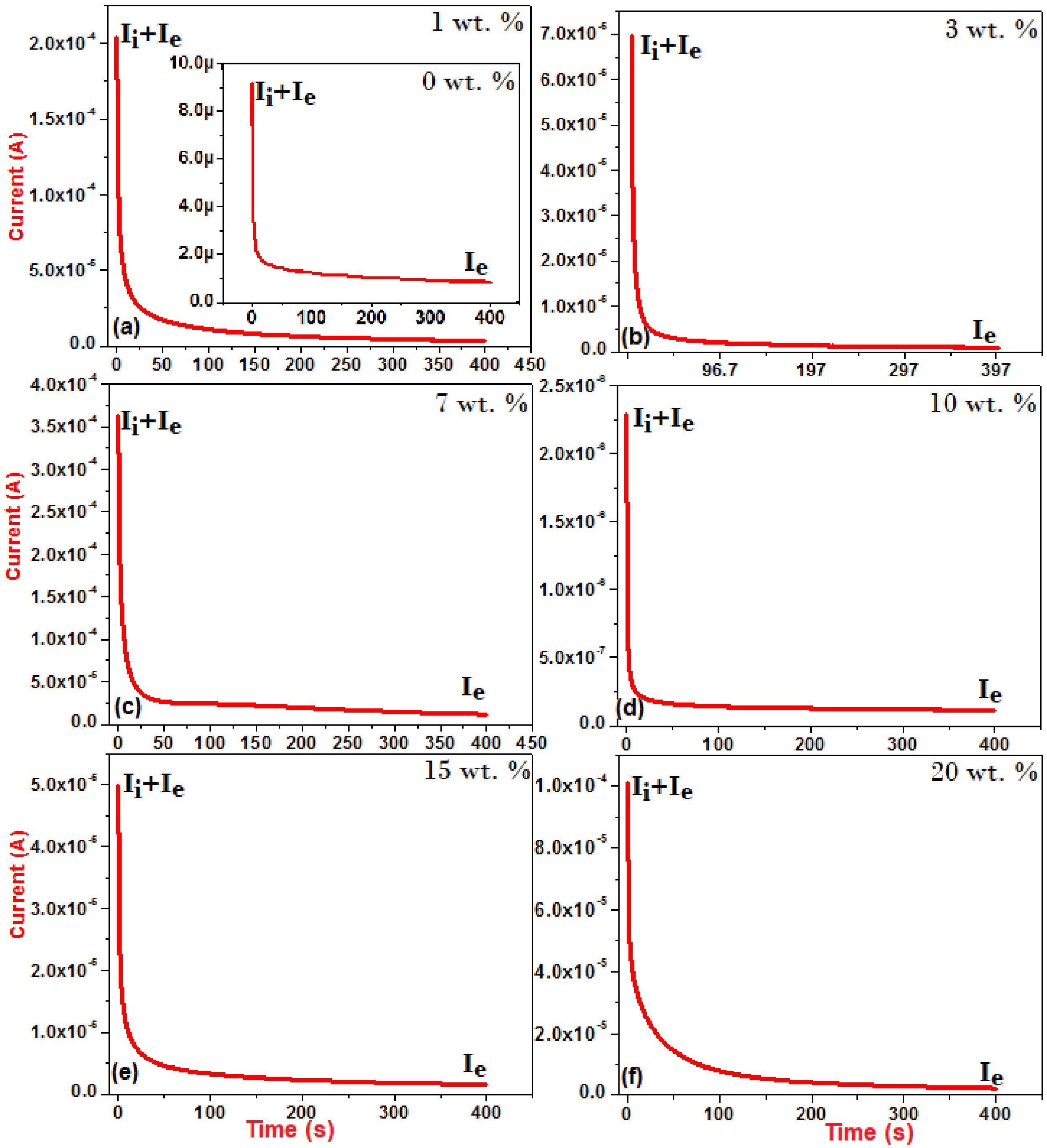

**Figure 14.** Ion transference number of SPE films (PEO-PVC)-LIPF$_6$+ *x* wt. % TiO$_2$ films **(a)** *x* =1 **(b)** x=3 **(c)** x=7 **(d)** x=10 **(e)** 15 **(f)** x=20 (x=0 wt. % in inset).

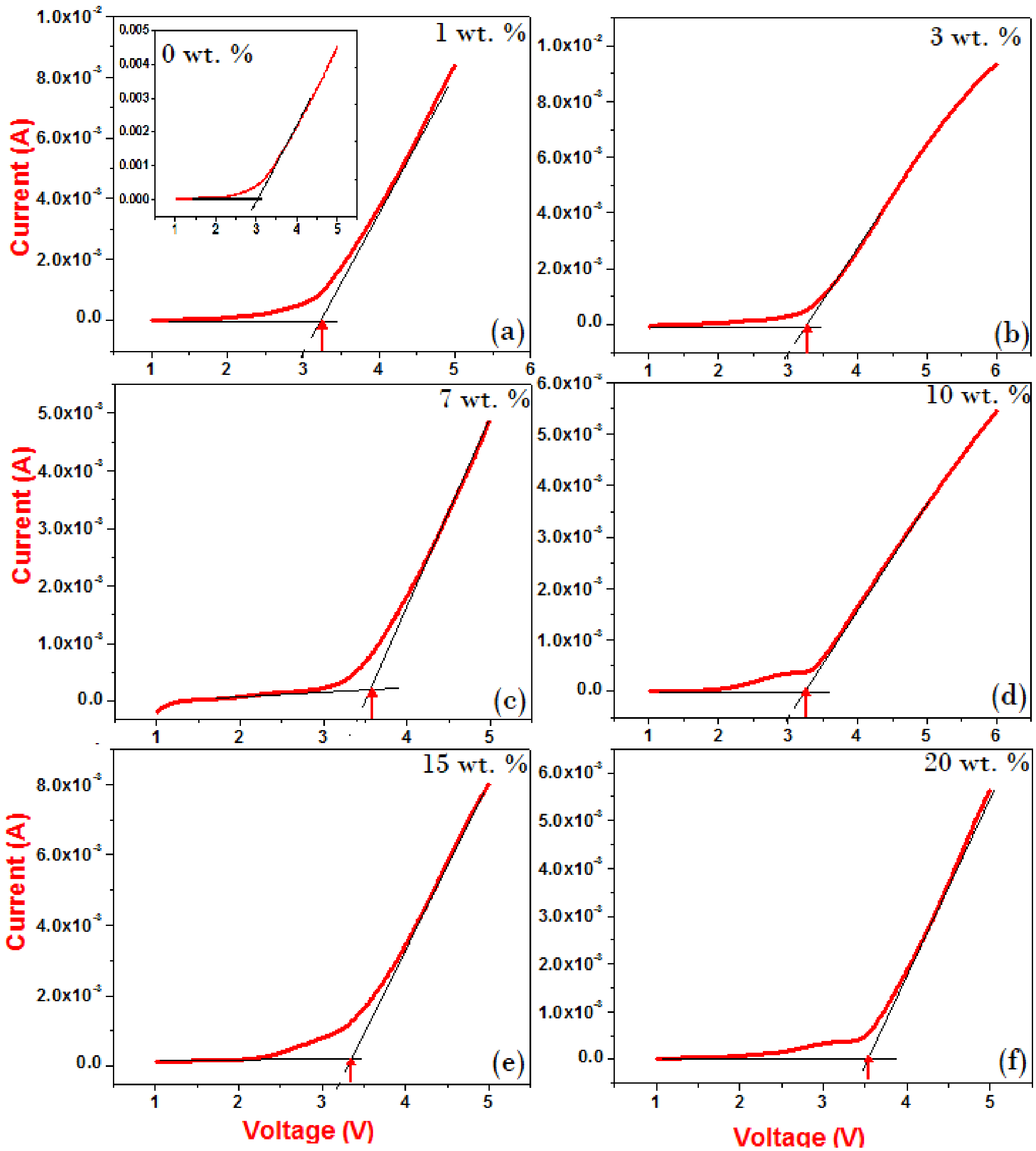

**Figure 15.** Linear Sweep Voltammetry of SPE films (PEO-PVC)-LIPF$_6$+ *x* wt. % TiO$_2$ films **(a)** *x* =1 **(b)** x=3 **(c)** x=7 **(d)** x=10 **(e)** 15 **(f)** x=20 (x=0 wt. % in inset).

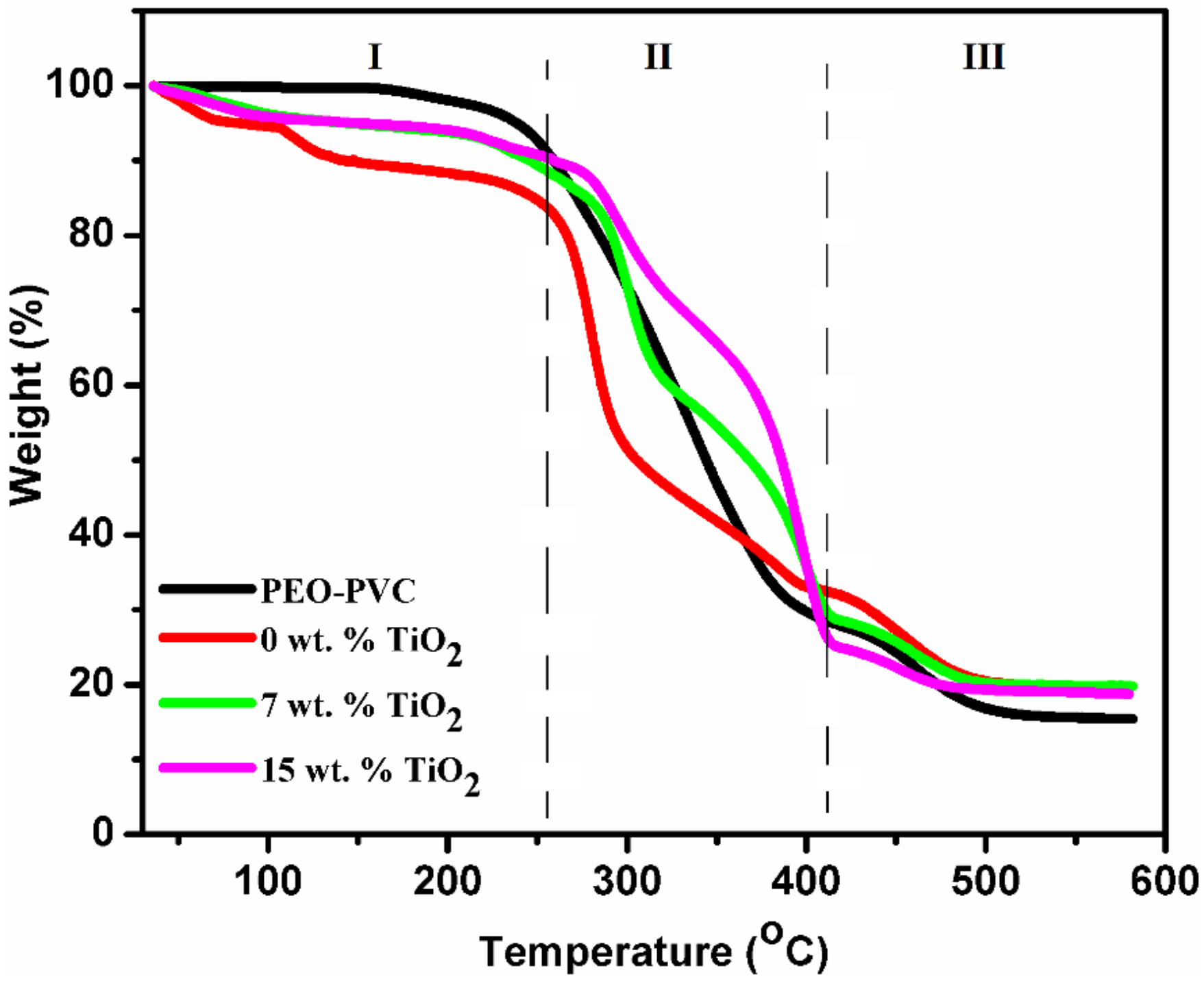

**Figure 16.** TGA curves of PEO-PVC blend, 0 wt. % $TiO_2$, 7 wt. % $TiO_2$ and 15 wt. % $TiO_2$.

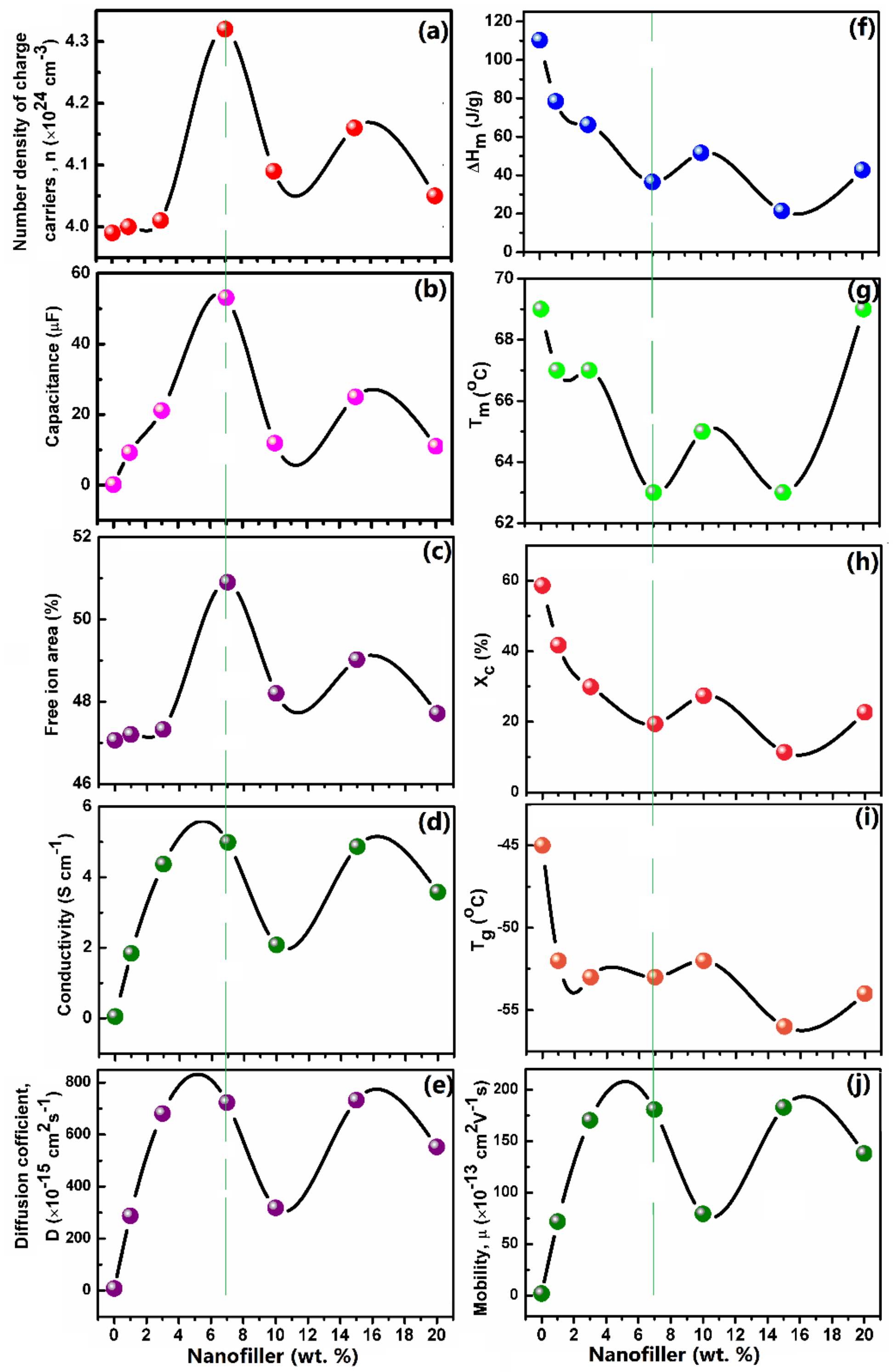

**Figure 17.** Plot of variation in **(a)** charge carrier number density n, **(b)**double layer capacitance $C_{dl}$, **(c)** free ions area (%), **(d)** dc conductivity σ, **(e)** diffusion coefficient D, **(f)** enthalpy of melting $\Delta H_m$ (J/g), **(g)** melting temperature $T_m$ (°C), **(h)** crystallinity $X_c$ (%) , **(i)** glass transition temperature $T_g$ (°C), and **(j)** charge carrier mobility (μ).

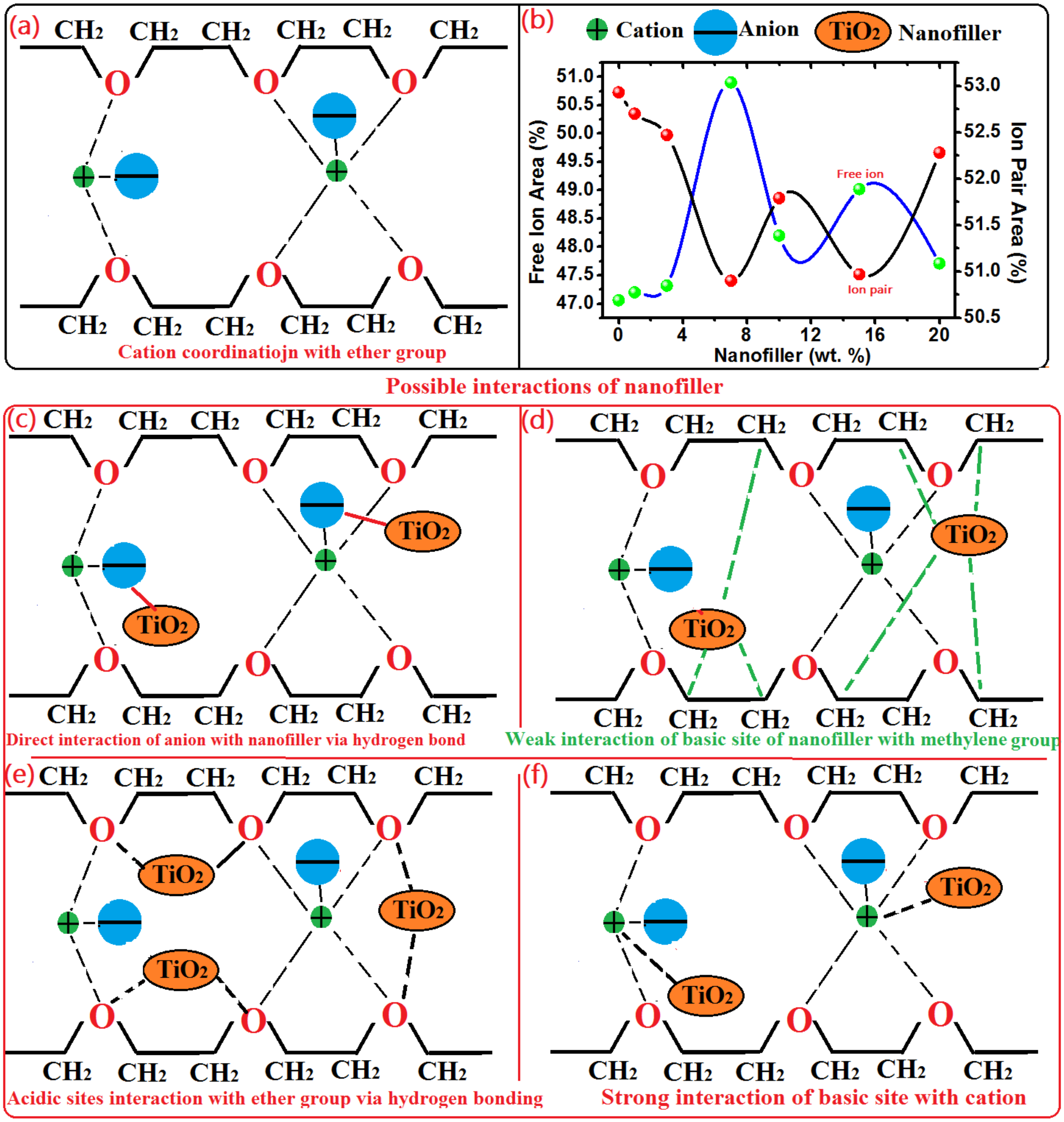

**Figure 18.** Various possible interactions between polymer-ion-nanofiller.

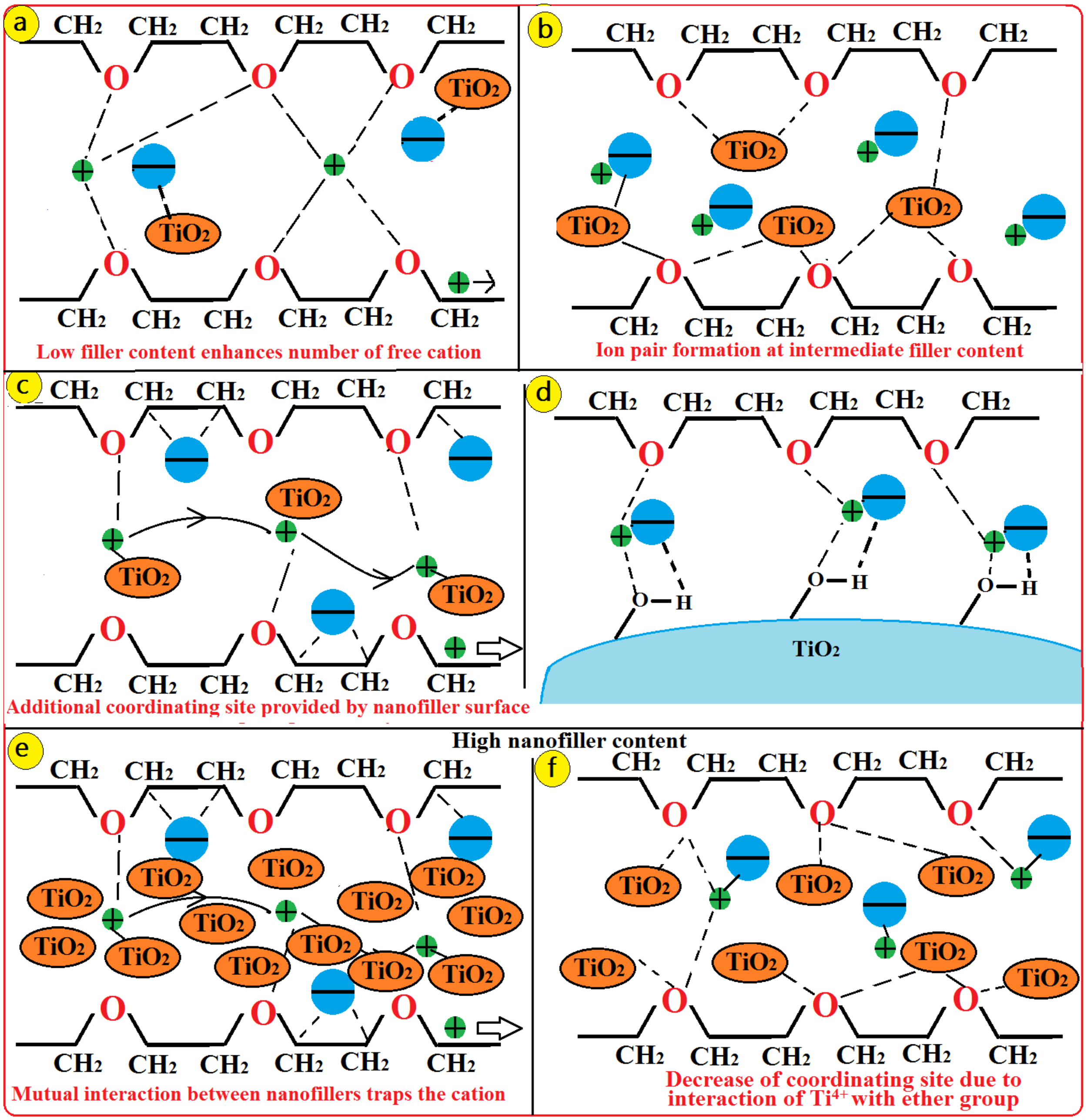

**Figure 19.** A pictorial model to illustrate proposed ion conduction mechanism for enhancement of ionic conductivity in solid polymer electrolytes on basis of nanofiller content.

**Table 1.** Values of Bragg's angle 2θ , Basal spacing (d), FWHM (β), crystallite size (L) and inter-chain separation (R) corresponding to 120 and 112,032 reflection planes of PEO in PEO-PVC+$LiPF_6$–$x$ wt. % $TiO_2$ nanocomposite films.

| Sample Code | 2θ (degree) | d (Å) | R (Å) | β ×$10^3$ (radians) | L (nm) |
|---|---|---|---|---|---|
| *PEO 120 reflection plane parameters* | | | | | |
| PEO-PVC Blend | 19.23 | 4.61 | 5.76 | 19.49 | 7.79 |
| x=0 | 18.45 | 4.80 | 6.00 | 13.75 | 10.60 |
| x=1 | 19.07 | 4.65 | 5.81 | 16.88 | 8.65 |
| x=3 | 19.16 | 4.62 | 5.78 | 12.88 | 11.33 |
| x=7 | 18.91 | 4.68 | 5.85 | 24.19 | 6.03 |
| x=10 | 18.98 | 4.67 | 5.86 | 20.18 | 7.23 |
| x=15 | 19.16 | 4.62 | 5.78 | 14.27 | 10.23 |
| x=20 | 19.10 | 4.64 | 5.79 | 16.01 | 9.12 |
| *PEO 112,032 reflection plane parameters* | | | | | |
| PEO-PVC Blend | 11.74 | 3.78 | 4.72 | 20.71 | 7.10 |
| x=0 | 11.59 | 3.83 | 4.78 | 37.24 | 3.94 |
| x=1 | 11.65 | 3.81 | 4.76 | 18.97 | 7.75 |
| x=3 | 11.61 | 3.82 | 4.78 | 16.18 | 9.08 |
| x=7 | 11.55 | 3.84 | 4.80 | 16.36 | 8.98 |
| x=10 | 11.56 | 3.84 | 4.80 | 18.27 | 8.04 |
| x=15 | 11.71 | 3.79 | 4.73 | 17.92 | 8.20 |
| x=20 | 11.74 | 3.78 | 4.72 | 22.97 | 6.40 |

**Table 2.** FTIR band identification and assignment in SPE films (PEO-PVC)-$LIPF_6$+ $x$ wt.% $TiO_2$ ($0 \leq x \leq 20$).

| **PEO-PVC** | **x =0** | **x =1** | **x =3** | **x =7** | **x =10** | **x =15** | **x =20** | **Band Assignment** | **Ref.** |
|---|---|---|---|---|---|---|---|---|---|
| 620 | 617 | 625 | 618 | 620 | 623 | 610 | 616 | (C-Cl) st. | [20] |
| 696 | 731 | 733 | 732 | 727 | 729 | 7729 | 690 | | |
| 844 | 844 | 841 | 842 | 846 | 842 | 842 | 842 | $CH_2$ Rocking/$PF_6^-$ | [50] |
| 951 | 956 | 956 | 958 | 960 | 958 | 962 | 958 | *trans* $(C\text{-}H)_w$ | [51] |
| 1114 | 1112 | 1112 | 1112 | 1097 | 1107 | 1122 | 1098 | $\nu(COC)_{s/a}$ | [4], [50] |
| 1240 | 1245 | 1244 | 1246 | 1247 | 1253 | 1259 | 1247 | $\text{CH Rocking}/\tau(CH_2)_s$ | [20] |
| 1350 | 1349 | 1345 | 1342 | 1338 | 1339 | 1343 | 1343 | $w(CH_2)_s$ | [36] |
| 1434 | 1426 | 1429 | 1432 | 1434 | 1435 | 1438 | 1429 | $\delta(CH_2)_s$ | [46] |
| 1467 | 1467 | 1463 | 1462 | 1466 | 1457 | 1458 | 1471 | $\delta(CH_2)_a$ | [46], [54] |
| 2888 | 2885 | 2884 | 2883 | 2895 | 2883 | 2885 | 2881 | $\nu(CH_2)_s$ | [5] |
| 2914 | 2910 | 2909 | 2903 | 2918 | 2919 | 2921 | 2911 | $\nu(CH_2)_a$ | [5] |

**Table 3.** The peak position of deconvoluted free ion and ion pair peak of SPE film based on (PEO-PVC)-$LIPF_6$+ $x$ wt. % $TiO_2$ ($0 \leq x \leq 20$).

| Nanofiller (wt. %) | Free Ion | | Ion Pair | | Correlation Coff. ($r^2$) |
|---|---|---|---|---|---|
| | ***Peak ($cm^{-1}$)*** | ***Area (%)*** | ***Peak ($cm^{-1}$)*** | ***Area (%)*** | |
| 0 | 835 | 47.06 | 847 | 52.93 | 0.98719980 |
| 3 | 836 | 47.32 | 847 | 52.47 | 0.99081045 |
| 7 | 846 | 50.90 | 846 | 50.90 | 0.98454013 |
| 10 | 835 | 48.20 | 847 | 51.79 | 0.99512452 |
| 15 | 835 | 49.02 | 847 | 50.97 | 0.97854570 |
| 20 | 836 | 47.71 | 847 | 52.28 | 0.99092205 |

**Table 4.** Glass transition temperature ($T_g$), melting temperature ($T_m$), enthalpy of melting and degree of crystallinity for polymer blend, polymer salt, and *x* wt. % (x=0, 1, 3, 7, 10, 15, 20) nanofiller added solid polymer electrolytes.

| Polymer Electrolyte | $T_g$(°C) | $T_m$ (°C) | $\Delta H_m$ (J/g) | $X_c$ (%) |
|---|---|---|---|---|
| PEO-PVC Blend | -30 | 78 | 174.8 | 92.97 |
| 0 | -45 | 69 | 110.2 | 58.61 |
| 1 | -52 | 67 | 78.42 | 41.71 |
| 3 | -53 | 67 | 66.28 | 29.93 |
| 7 | -53 | 63 | 36.48 | 19.40 |
| 10 | -52 | 65 | 51.56 | 27.42 |
| 15 | -56 | 63 | 21.37 | 11.36 |
| 20 | -54 | 69 | 42.7 | 22.71 |

**Table 5.** Fitted Parameters of Nonlinear Least Squares (NLS) Fit of the samples comprising of SPE films (PEO-PVC)-$LIPF_6$+ $x$ wt. % $TiO_2$ ($0 \leq x \leq 20$).

| **Sample Name(x wt. % $TiO_2$)** | **$Q_1$** | **$n_1$** | **$Q_2$** | **$n_2$** | **$C_{dl}$(μF)** |
|---|---|---|---|---|---|
| PEO-PVC Blend | $3.5\times10^{-4}$ | 0.052 | $5.42\times10^{-5}$ | 0.67 | 0.10 |
| x=0 | $1.5\times10^{-3}$ | 0.035 | $2.18\times10^{-5}$ | 0.66 | 0.18 |
| x=1 | $4.3\times10^{-4}$ | 0.018 | $4.31\times10^{-5}$ | 0.74 | 9.21 |
| x=3 | $2.72\times10^{-5}$ | 0.015 | $3.79\times10^{-7}$ | 0.82 | 21.1 |
| x=7 | $2.8\times10^{-3}$ | 0.002 | $6.57\times10^{-5}$ | 0.8 | 53.1 |
| x=10 | $1.73\times10^{-5}$ | 0.045 | $1.86\times10^{-5}$ | 0.76 | 11.9 |
| x=15 | $4.3\times10^{-3}$ | 0.025 | $3.97\times10^{-5}$ | 0.8 | 25 |
| x=20 | $1.79\times10^{-5}$ | 0.037 | $2.74\times10^{-3}$ | 0.72 | 11 |

**Table 6.** Different contributions of electrical conductivity and transport number for (PEO-PVC)-$LIPF_6$+ $x$ wt. % $TiO_2$ ($0 \leq x \leq 20$).

| Sample Name (x wt. % $TiO_2$) | Transference Number | Electrical Conductivity (S $cm^{-1}$) | Electronic Conductivity (S $cm^{-1}$) | Ionic Conductivity (S $cm^{-1}$) |
|---|---|---|---|---|
| PEO-PVC Blend | 0.67 | $3.33\times10^{-8}$ | $1.09\times10^{-8}$ | $2.23\times10^{-8}$ |
| x=0 | 0.90 | $0.51\times10^{-6}$ | $5\times10^{-10}$ | $0.45\times10^{-6}$ |
| x=1 | 0.99 | $1.84\times10^{-5}$ | $1.8\times10^{-10}$ | $1.82\times10^{-5}$ |
| x=3 | 0.98 | $4.37\times10^{-5}$ | $8.7\times10^{-10}$ | $4.28\times10^{-5}$ |
| x=7 | 0.96 | $4.99\times10^{-5}$ | $1.99\times10^{-9}$ | $4.89\times10^{-5}$ |
| x=10 | 0.95 | $2.08\times10^{-5}$ | $1.04\times10^{-9}$ | $1.97\times10^{-5}$ |
| x=15 | 0.97 | $4.87\times10^{-5}$ | $1.46\times10^{-9}$ | $4.72\times10^{-5}$ |
| x=20 | 0.98 | $3.58\times10^{-5}$ | $7.1\times10^{-10}$ | $3.50\times10^{-5}$ |

**Table 7**. The values of $V_{Total}$, free ions (%), n, $\mu$, D obtained using the FTIR method.

| Nanofiller Content (wt. %) | $V_{Total}$ ($\times 10^{-1}\ cm^3$) | Free Ions Area (%) | n ($\times 10^{24}\ cm^{-3}$) | $\mu$ ($\times 10^{-13}\ cm^2V^{-1}s$) | D ($\times 10^{-15}\ cm^2s^{-1}$) |
|---|---|---|---|---|---|
| 0 | 0.2 | 47.06 | 3.99 | 7.98 | 1.99 |
| 1 | 0.2 | 47.20 | 4.00 | 287.5 | 71.87 |
| 3 | 0.2 | 47.32 | 4.01 | 681.10 | 170.27 |
| 7 | 0.2 | 50.90 | 4.32 | 723.37 | 180.84 |
| 10 | 0.2 | 48.20 | 4.09 | 317.84 | 79.46 |
| 15 | 0.2 | 49.02 | 4.16 | 731.67 | 182.91 |
| 20 | 0.2 | 47.71 | 4.05 | 552.46 | 138.11 |